\documentclass[aps,prb,twocolumn,groupedaddress,superscriptaddress]{revtex4}

\usepackage{amsmath}
\usepackage{graphicx}
\usepackage{subfigure}
\usepackage{color}

\begin{document}


\title{Signatures of strong correlation effects in RIXS on Cuprates}
\author{Wan-Ju Li}
\affiliation{Institute of Physics, Academia Sinica, Taipei 11529, Taiwan}
\author{Cheng-Ju Lin}
\affiliation{Institute of Physics, Academia Sinica, Taipei 11529, Taiwan}
\affiliation{Department of Physics and Institute for Quantum Information and Matter,
California Institute of Technology, Pasadena, CA 91125, USA}
\author{Ting-Kuo Lee}
\affiliation{Institute of Physics, Academia Sinica, Taipei 11529, Taiwan}


\author{}

\affiliation{}


\date{\today}

\begin{abstract}
Recently, spin excitations in doped cuprates are measured using the resonant inelastic X-ray scattering (RIXS). The paramagnon dispersions show the large hardening effect in the electron-doped systems and seemingly doping-independence in the hole-doped systems, with the energy scales comparable to that of the antiferromagnetic  magnons. This anomalous hardening effect was partially explained by using the strong coupling $t-J$ model but with a three-site term(Nature communications 5, 3314 (2014)). However we show that hardening effect is  a signature of strong coupling physics even without including this extra term.  By considering the $t-t'-t''-J$ model and using the Slave-Boson (SB) mean field theory, we obtain, via the spin-spin susceptibility, the spin excitations in qualitative agreement with the experiments. These anomalies  is mainly due to the doping-dependent bandwidth. We further discuss the interplay between  particle-hole-like and paramagnon-like excitations in the RIXS measurements.
\end{abstract}


\maketitle
\section{Introduction}


It is generally believed that magnetic interaction may be responsible for the superconductivity in cuprates\cite{Rev2006}. Recently, the development of the resonant inelastic X-ray scattering (RIXS)\cite{Ament2011,Dean2015} enables experimentalists to measure the spin excitations over a more comprehensive region of the Brillouin zone than the conventional inelastic neutron scattering (INS) experiments\cite{Le-Tacon:2011fk}. A large family of both electron- and hole-doped materials have been investigated and the spin excitations are reported to resemble the dispersion of the antiferromagnetic (AFM) magnon in the paramagnetic phase, called paramagnon.\cite{Le-Tacon:2011fk,LeTacon2013,Dean2013,Braicovich2010,Lee2014,Wakimoto2015,Minola2015,Guarise2014,Dean2014,Dean2013b} While many publications demonstrate the magnetic nature of the paramagnon\cite{Le-Tacon:2011fk,LeTacon2013,Dean2013,Minola2015}, it is argued that the analogy with spin waves is only partial\cite{Guarise2014} and the itinerant nature of this magnetic excitation cannot be ignored\cite{Wakimoto2015,Dean2014,Monney2016,Huang2016,Roland2013}. The strong flavor of AFM magnon also reinvigorates an old debate  that may be the AFM fluctuations, seemingly much more robust, is more important for cuprates’  superconductivity than strong correlation just as for iron-based superconductors \cite{Si2016}. 

In addition to the magnetic and itinerant nature of this spin excitation, the anomalous doping dependence of the energy dispersions is very intriguing. Contrary to the notion suggested by the INS experiments, the paramagnon dispersions measured by RIXS are of the similar excitation energy scale among different hole dopings\cite{Dean2013b}. Moreover, the paramagnons show the anomalously large hardening of the energy dispersions in the electron-doped cuprates\cite{Lee2014}, while hole-doped cuprates do not exhibit much softening as hole concentration increases. This is contrary to the expectation that paramagnon dispersion  will soften when there are more itinerant carriers involved in screening. Theoretically, Jia et al.,\cite{Jia2014} study an effective single-band Hubbard model using the determinant Quantum Monte Carlo (DQMC) and obtain results consistent with experiments. To explain the physics of the hardening effect in e-doped systems, they introduce an 3-site exchange term in t-J type model. Although their exact diagonalization (ED) calculations including the 3-site exchange term do reproduce the correct scale of the hardening effect, their results also show that the hardening effect appears even before their introduction of this extra term. This indicates that the hardening effect is intrinsic in the strong-correlation picture of t-J type model without adding any extra interaction terms. In this work, we would like to point out that these anomalies are signatures of the strong correlation. More precisely, the Mott physics provides a strong bandwidth renormalization as shown by using Gutzwiller approximation to treat the constraint of no doubly occupied sites in the t-J model\cite{Zhang:1988uq}. To illustrate this idea in the simplest possible way, we shall use Slave-Boson theory\cite{Lee1992,Bickers1987} to include the strong correlation effect.  Investigating the model in AFM, paramagnetic (PM), and superconducting (SC) phases, we calculate the spin-spin susceptibility and recognize that it is the enhancement  of the bandwidth with the dopant density that hardens the energy dispersion, a result of Mott physics accounting for the anomalous experimental observations. Furthermore, based on our calculations, we argue that the experimentally-observed spin excitations are mixtures of both particle-hole-like and paramagnon-like excitations. It is noted that a recent work \cite{Roland2013} applied similar methods to the calculations of the Raman spectra of doped cuprates and their results are consistent with our calculations.

\section{Theoretical model}
\subsection{AFM and PM phases}
The $t-t'-t''-J$ model Hamiltonian is written as

\begin{align} \label{t-t'-t''-J}
H = &-t\sum_ {<ij>\sigma} (c_{i \sigma}^{\dagger}c_{j\sigma}+\text{h.c.})
-t'\sum_{<ij>_2 \sigma} (c_{i \sigma}^{\dagger}c_{j \sigma}+\text{h.c.})  \notag  \\
& -t''\sum_{<ij>_3\sigma} (c_{i\sigma}^{\dagger}c_{j\sigma}+\text{h.c.})
+J\sum_{<ij>\sigma} (\mathbf{S}_i\cdot \mathbf{S}_j-\frac14 n_in_j)  \notag \\
& -\mu_0\sum_{i\sigma}c^{\dagger}_{i\sigma}c_{i\sigma},
\end{align}

where $<>$,$<>_2$, and $<>_3$ represent the nearest neighbour (n.n.), second n.n., and third n.n., respectively. In the presence of strong Coulomb repulsion, each site is at most singly occupied.  We treat the Hamiltonian by the Slave-Boson mean-field theory\cite{Yuan2004,Yuan2005}, i.e., $c_{i\sigma}=b^{\dagger}_if_{i\sigma}$ and $\mathbf{S}_i=\frac{1}{2}\sum_{\sigma \sigma '}f^{\dagger}_{i \sigma} \boldsymbol{\tau}^{\sigma \sigma '}f_{i\sigma '}$, where $\boldsymbol{\tau}^{\sigma \sigma '}$ are the Pauli matrices. Taking the mean-field parameters as $m=(-1)^i\langle S^z_i\rangle$, the AFM order, and $X=\langle f^\dagger_{i\sigma}f_{j\sigma} \rangle $, the uniform hopping term, the Hamiltonian is written in the momentum space with bosonic operators being replaced by the square root of the average hole density

\begin{align} \label{MFH}
H=&\sum_{k,\sigma}^{ } {'}(\epsilon_kf^{\dagger}_{k\sigma}f_{k\sigma}+\epsilon_{k+Q}f^{\dagger}_{k+Q\sigma}f_{k+Q\sigma}) \notag \\
&-2Jm\sum_{k,\sigma}^{ } {'}\sigma(f^{\dagger}_{k\sigma}f_{k+Q\sigma}+\text{h.c.})+2NJ(X^2+m^2),
\end{align}

where $\sum_{k}^{ }{'} $ indicates the summation is over the magnetic Brillouin zone (MBZ): $-\pi < k_x \pm k_y \leq \pi$, $\epsilon_k=(-2t\delta-JX)(\cos k_x+\cos k_y)-4t'\delta\cos k_x \cos k_y-2t''\delta(\cos 2k_x + \cos 2k_y)-\mu$, $Q=(\pi,\pi)$, and $N$ is the total number of lattice sites. Due to the strong correlation between electrons, the hopping terms of the energy bands are modulated by the dopant density, which is similar to the Gutzwiller approximation to  replace the constraint of forbidding double occupancy with a renormalization factor, $g_t=2\delta/(1+\delta)$\cite{Zhang:1988uq}. Here our factor is about two times smaller, we will discuss this below.

By taking the unitary transformations: $f_{k\sigma}=\cos \theta_k \alpha_{k\sigma}+\sigma\sin\theta_{k}\beta_{k\sigma}$ and $f_{k+Q\sigma}=-\sigma\sin \theta_k \alpha_{k\sigma}+\cos\theta_{k}\beta_{k\sigma}$ with $\cos2\theta_k=(\epsilon_{k+Q}-\epsilon_{k})/\gamma_k$, $\sin2\theta_k=-4Jm/\gamma_k$, and $\gamma_k=\sqrt{(\epsilon_{k+Q}-\epsilon_{k})^2+(4Jm)^2}$
, we obtain
\begin{equation}
H=\sum_{k,\sigma}^{ } {'}(\xi_{k\alpha}\alpha^{\dagger}_{k\sigma}\alpha_{k\sigma}+\xi_{k\beta}\beta^{\dagger}_{k\sigma}\beta_{k\sigma})+2NJ(X^2+m^2),
\end{equation}
with the energy bands $\xi_{k\alpha,\beta}=(\epsilon_k+\epsilon_{k+Q}\mp \gamma_k)/2 $. The free energy is given by $F=-2T\sum_{\eta=\alpha,\beta}^{ } \sum_{k}^{ } {'} \ln(1+e^{-\xi_{k \eta} / T})+2NJ(X^2+m^2)$. The mean-field parameters $m$ and $X$, as well as the chemical potential $\mu$ are computed self-consistently by the conditions $\partial F/\partial m=0$, $\partial F/ \partial X=0$ and $-(\partial F/\partial \mu)=N(1-\delta)$ at zero temperature. The model parameters taken through out the work are $t=1.0$, $t'=-0.3$, $t''=0.2$ and $J=0.3$ for the hole-doped case; while $t'=0.3$ and $t''=-0.2$ for the electron-doped case\cite{TK2003}.

The transverse spin susceptibility is defined as
\begin{equation}
\chi^{\pm}_{(0)}(q,q',\tau)=\frac{1}{N}\langle T_\tau S^+_q(\tau)S^-_{-q'}(0) \rangle_{(0)} ,
\end{equation}
where $\langle \cdots \rangle_0$ means the thermal average on the eigenstates of the mean-field Hamiltonian, $S^+_q=\sum_i S^+_i e^{iq\cdot R_i}$ and $S^-_q=(S^+_{-q})^\dagger$. The residual fluctuations of the spin-spin interaction is taken into account by the Random Phase Approximation (RPA).\cite{Yuan2005}

Because of the non-vanishing off-diagonal correlation function as a result of antiferromagnetism, the spin susceptibility is written as a matrix
\begin{equation}
\label{AFMRPAchi}
 \hat{\chi}^{\pm} =  \left( \begin{array}{cc}
\chi^\pm(q,i\omega_n) & \chi^\pm(q,q+Q,i\omega_n)  \\
\chi^\pm(q+Q,q,i\omega_n) & \chi^\pm(q+Q,i\omega_n)   \end{array} \right).
\end{equation}
The diagonal term is given as
\begin{align}
\chi^\pm_0(q,i\omega_n)= &-\frac{1}{N}\sum_k^{}{'}[\cos^2(\theta_k+\theta_{k+q})(F_{\alpha\alpha}+F_{\beta\beta})  \notag \\
&+\sin^2(\theta_k+\theta_{k+q})(F_{\alpha\beta}+F_{\beta\alpha})],
\end{align}
and the off-diagonal term as

\begin{align}
\chi^\pm_0(q,q+Q,i\omega_n)= &\frac{1}{2N}\sum_k^{}{'}[(\sin 2\theta_{k+q}-\sin 2\theta_{k})(F_{\alpha\alpha}-F_{\beta\beta}) \notag \\
&+(\sin 2\theta_{k+q}+\sin 2\theta_{k})(F_{\alpha\beta}-F_{\beta\alpha})],
\end{align}
with the abbreviations
\begin{align*}
F_{\eta\eta'}=\frac{n(\xi_{k+q,\eta})-n(\xi_{k,\eta'})}{i\omega_n+\xi_{k+q,\eta}-\xi_{k,\eta'}}\; (\eta,\eta'=\alpha,\beta).
\end{align*}
$n(z)=1/(1+e^{(z/T)})$ is the Fermi function, and $i\omega_n$ are the Matsubara frequencies. The RPA result is given as
\begin{equation}
\label{RPAAFM}
\hat{\chi}^\pm_{\text{RPA}}=\hat{\chi}_0^\pm[I+\hat{\chi}^\pm_0 \hat{J}]^{-1},
\end{equation}
where $I$ is the identity matrix and
\begin{equation}
\hat{J}= \left( \begin{array}{cc}
J(q) & 0 \\
0 & J(q+Q)
\end{array} \right ),
\end{equation}
with $J(q)=J(\cos q_x+\cos q_y)$. We can see from the equations above that there are two parts contributing to the spin-spin excitations. One is the particle-hole excitation constituted of the interband ($\alpha$ to $\beta$ or $\beta$ to $\alpha$) and the intraband ($\alpha$ to $\alpha$ or $\beta$ to $\beta$) excitations, which are described by $\chi_0^\pm$, or more specifically, the term $F_{\eta\eta'}$. The other is the collective spin-wave excitation mode, which is a result of the additional poles generated by the RPA calculation from $\det[I+\hat{\chi}^\pm_0\hat{J}]=0$.

For the paramagnetic case, the off-diagonal terms of the spin susceptibility vanishes and the result is simply given as
\begin{equation}
\chi_0^\pm(q,i\omega_n)=-\frac{1}{N}\sum_k^{ }\frac{n(\epsilon _{k+q})-n(\epsilon _{k})}{i\omega_n+\epsilon_{k+q}-\epsilon_k},
\end{equation}
and
\begin{equation}\label{RPANoAFM}
\chi_{\text{RPA}}^\pm(q,i\omega_n)=\frac{\chi_0^\pm(q,i\omega_n)}{1+\chi_0^\pm(q,i\omega_n)J(q)}.
\end{equation}
In this case, usually the denominator does not have a pole and the numerator with particle-hole excitation becomes dominant. The information of the excitations is contained in $\chi^\pm_0$, and the weights are modified by the denominator in the RPA calculation.

\subsection{Superconducting phase}
In the SC phase\cite{Li2000,Li2001,Li2002,Li2003,Patrick1992,Patrick1999,TK1997}, the mean-field Hamiltonian is obtained by decoupling the spin-spin interaction term $\mathbf{S}_i\cdot \mathbf{S}_j$ into pairing and direct hopping terms,\cite{Patrick1992} and choosing the mean-field parameters $\Delta_{ij}=\langle f_{i\uparrow}f_{j\downarrow}-f_{i\downarrow}f_{j\uparrow}\rangle=\pm\Delta_0$, $X_0=\sum_{\sigma} \langle f^{\dagger}_{i\sigma}f_{j\sigma}\rangle$ and $\langle b_i \rangle=\sqrt{\delta}$,

\begin{equation} \label{SCMFH}
H=\sum_{k,\sigma}^{ }\xi_kf^{\dagger}_{k\sigma}f_{k\sigma} -\sum_{k}^{ }\Delta_k(f^{\dagger}_{k\uparrow}f^{\dagger}_{-k\downarrow}+\text{h.c.})+2NJ’(X_0^2+\Delta_0^2),
\end{equation}
where $\xi_k=(-2\delta t-2J’X_0)(\cos k_x+\cos k_y)-4\delta t'\cos k_x\cos k_y-2\delta t''(\cos 2k_x+\cos 2k_y)-\mu$ and $\Delta_k=2J'\Delta_0(\cos k_x-\cos k_y)$, with $J'=3J/8$. Here $X_0$ includes hoppings of both spins.

The spin-spin susceptibility applying RPA is given by Eq.~(\ref{RPANoAFM}),with the numerator
\begin{align}
\chi_0^{\pm}(q,i\omega_n)&=-\frac{1}{N}\sum_{k;\eta, \eta'=\pm }{C_{\eta\eta'}}\frac{n(\eta E_{k+q})-n(\eta'E_{k})}{i\omega_n+\eta E_{k+q}-\eta'E_{k}}.,
\end{align}
where $E_k=\sqrt{\xi_k^2+\Delta_k^2}$  is the quasiparticle excitation energy in the SC state. The coefficients $C_{\eta\eta'}$ are given as following: $C_{++}=v_{k+q}^2v_k^2+u_{k+q}v_{k+q}u_kv_k$, $C_{--}=u_{k+q}^2u_k^2+u_{k+q}v_{k+q}u_kv_k$, $C_{+-}=v_{k+q}^2u_k^2-u_{k+q}v_{k+q}u_kv_k$ and $C_{-+}=u_{k+q}^2v_k^2-u_{k+q}v_{k+q}u_kv_k$, with the coherence factors $v_k^2=\frac{1}{2}(1-\frac{\xi_k}{E_k})$, $u_k^2=\frac{1}{2}(1+\frac{\xi_k}{E_k})$ and $u_kv_k=\frac{\Delta_k}{2E_k}$.

\section{results and comparison with experiments}
\begin{figure}
\includegraphics[scale=0.5]{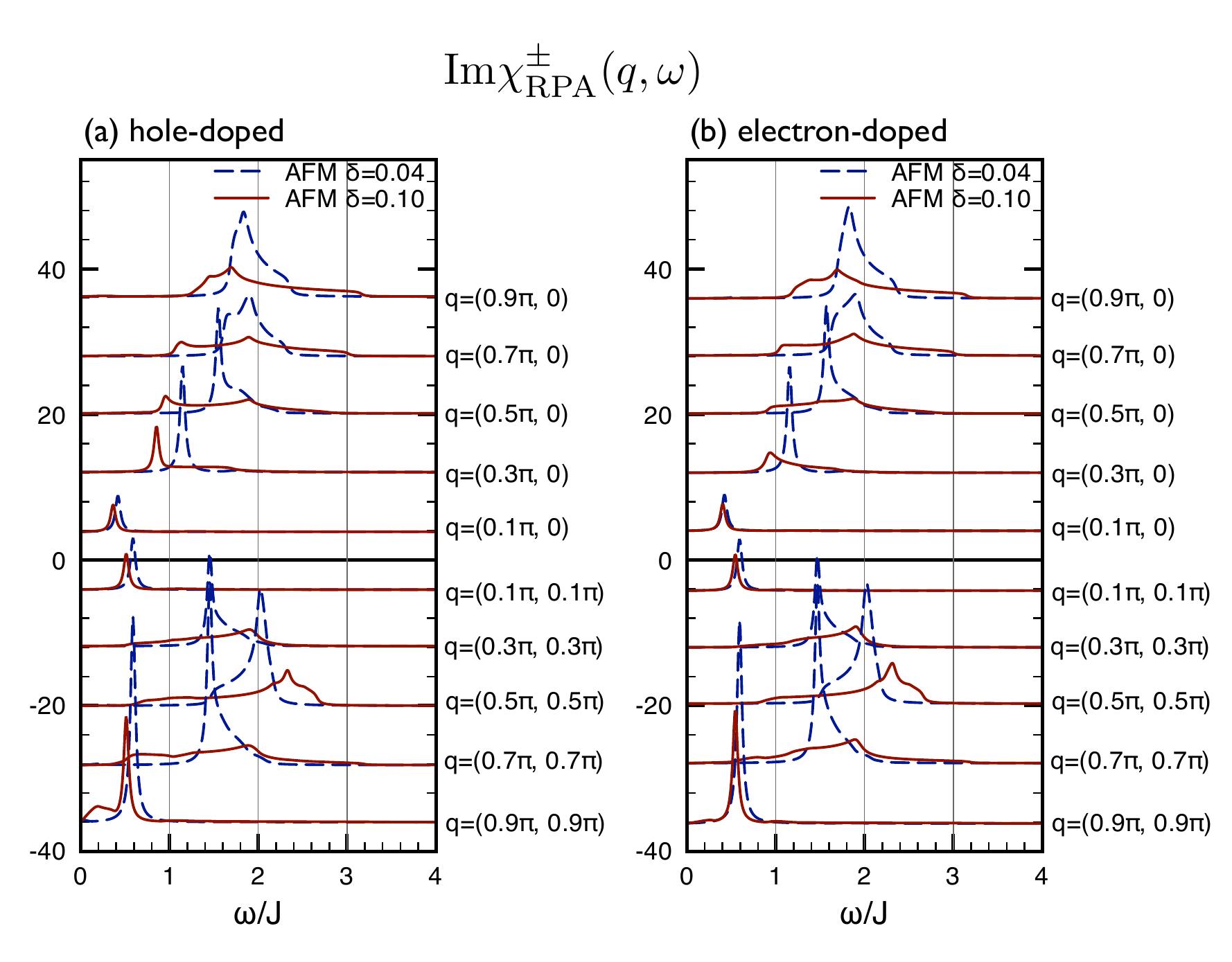}
\caption{\label{AFMRPA}The imaginary part of the spin-spin
susceptibility in the AFM phase along different momentum paths of (a)
hole-doped and (b) electron-doped cases. The dashed lines and the
solid lines represent $\delta=0.04$ and $\delta=0.10$ respectively.
The values in both cases of $q=(0.9, 0.9)\pi$
are reduced by 10; while $q=(0.7, 0.7)\pi$ and $q=(0.5, 0.5)\pi$ by
2, in order to fit into the figure. }
\end{figure}

The imaginary part of the spin susceptibility reflects the possible excitations, identified by the peaks, and their weights. In Fig.\ref{AFMRPA}, we show the imaginary part of $\chi^\pm_{\text{RPA}}$ in AFM cases, along different paths in the momentum space with analytical continuation $i\omega_n \rightarrow \omega+i\Gamma$ performed. We take the damping parameter $\Gamma=0.01|t|$ and $500\times500$ k points in the first Brillouin zone (or half of $500\times500$ k points in the MBZ) in all of our calculations. The energy spread of the excitation is related to the bandwidth, which is roughly proportional to the dopant density. Accordingly, the excitation spectrum broadens as the dopant density increases. Also, as the momentum $q$ gets larger along the $(\pi,0)$ direction (the upper panels in Fig. \ref{AFMRPA}) or gets closer to $\pi/2$ along the $(\pi,\pi)$ direction (the lower panels in Fig. \ref{AFMRPA}), the spectrum broaden as a result of the broader range of the accessible particle-hole excitation energies. This is an example showing that, in addition to the spin-wave excitations, particle-hole excitations also contribute to the spin susceptibility away from the resonant k-points. The detailed situation depends on the dopant density, as described above, but this general feature of the mixing of two types of excitations is prevailing throughout this work. We will discuss this below. 

\begin{figure}[h]
\centering
\subfigure[~susceptibility along $(\pi,0)$ direction]{
\includegraphics[width=.45\columnwidth]{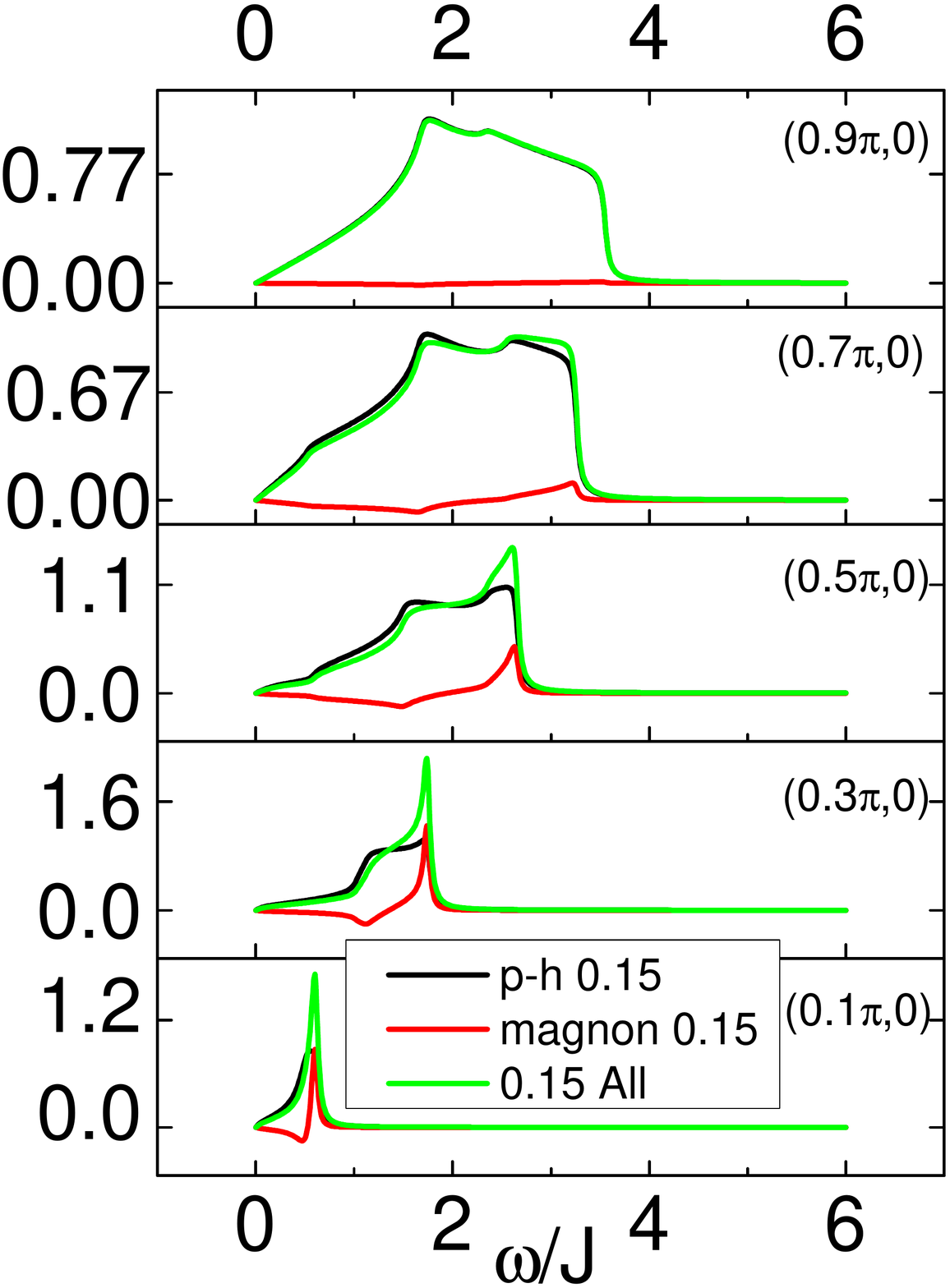}
\label{pi0} }
\subfigure[~susceptibility along $(\pi,\pi)$ direction]{
\includegraphics[width=.45\columnwidth]{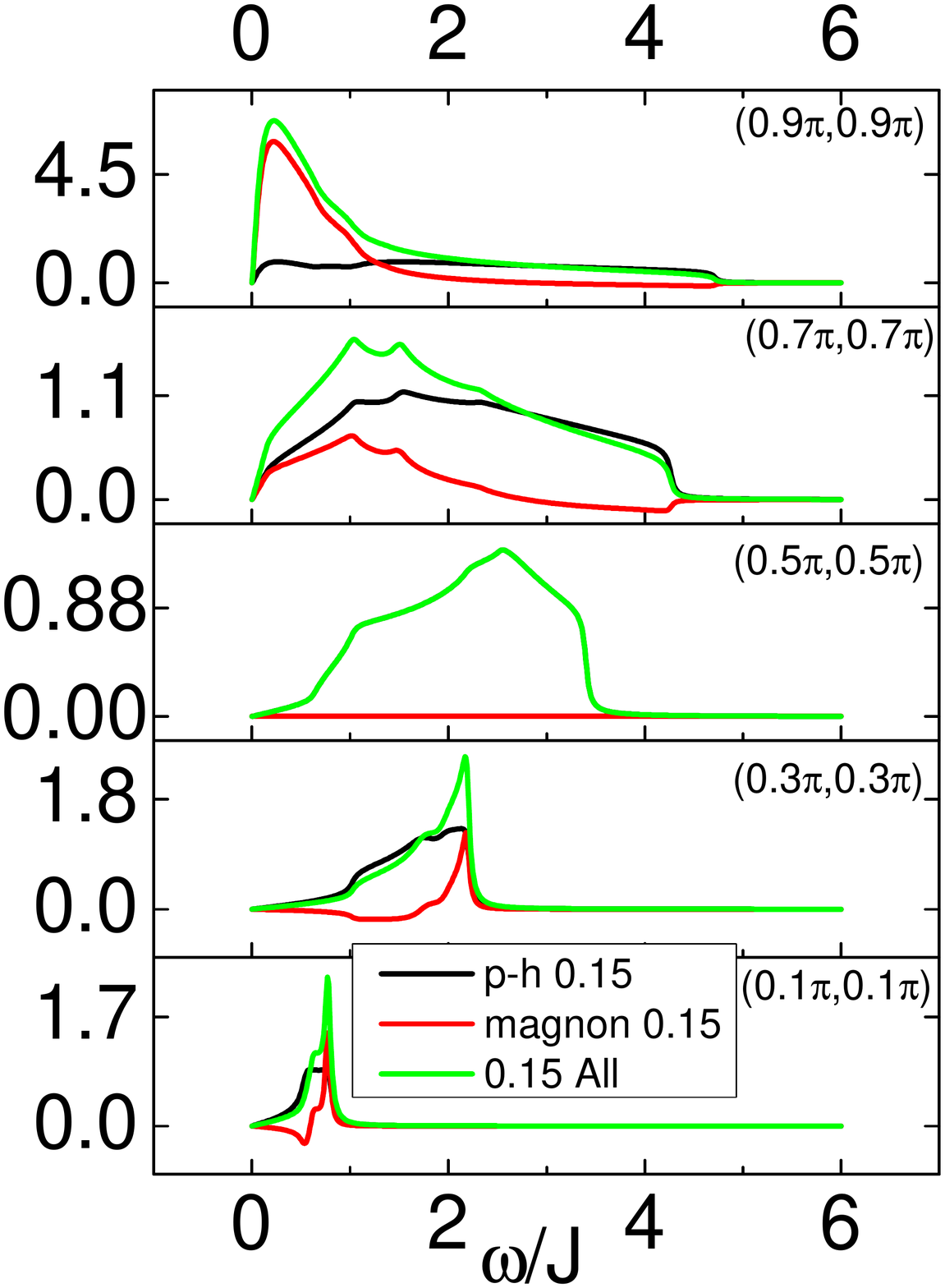}
\label{pipi} }
\caption{The spin susceptibility for e-doped system with doping = 0.15.(a) Along $(\pi,0)$ direction. (b) Along $(\pi,\pi)$ direction. At low-q and at around AFM point $(\pi,\pi)$, the peak of the total susceptibility (green) is mostly contributed by the spin-wave like excitations (red). For other momentum transfer $q$, the feature of the total susceptibility is dominated by the particle-hole like excitations (black).}
\label{chi0andchi}
\end{figure}

From Eqs. \ref{RPAAFM} and \ref{RPANoAFM}, the particle-hole like excitations appear when the numerators dominate while the spin-wave-like excitations show up for the divergence of the RPA-modified denominator. In the low-doping (AFM) regime, the resonance from the denominator is sharp and definite while in the PM phase, the resonance from the denominator become smooth and the contributions from the numerator becomes significant. In Fig.\ref{chi0andchi}, the spin susceptibility of an e-doped system with doping = 0.15 is calculated along both $(\pi,0)$ and $(\pi,\pi)$ directions. The total susceptibility (green lines) can be decomposed into contributions from the denominator (red lines) and the numerator (black lines). Along both directions, the denominator dominates at low-q and at around AFM point $(\pi,\pi)$ while the numerator has larger contributions at other momentum transfer. In comparison with Fig.\ref{AFMRPA}, larger dopant density corresponds to smaller $q$-range where spin-wave excitations dominate. However, the general pattern of two types of excitations is similar for all dopings.

Recent RIXS experiments seems to conclude that there are at least two distinct elementary excitations existing in the cuprate superconductors and their appearance depends both on the polarization of the incident photons and on the scattering geometry\cite{Minola2015,Huang2016}. One is the particle-hole like excitation whose resonance peaks change their positions with different incident photon energies. The other one is the spin-wave-like excitation (or, paramagnon) whose resonance peaks are located independently of the incident photon energy. However, from our calculations we find  these two excitations all contribute to the spin susceptibility. They may have a  dispersion similar as AFM spin waves at small momenta $q$, but they are mixed together.  Recent experiments\cite{Wakimoto2015,Huang2016} reporting the similar excitation energy scales measured using different photon polarizations and scattering geometries support this viewpoint.

\begin{figure}
 \centering
\subfigure[~dispersion for hole-doped cases]{
\includegraphics[width=.9\columnwidth]{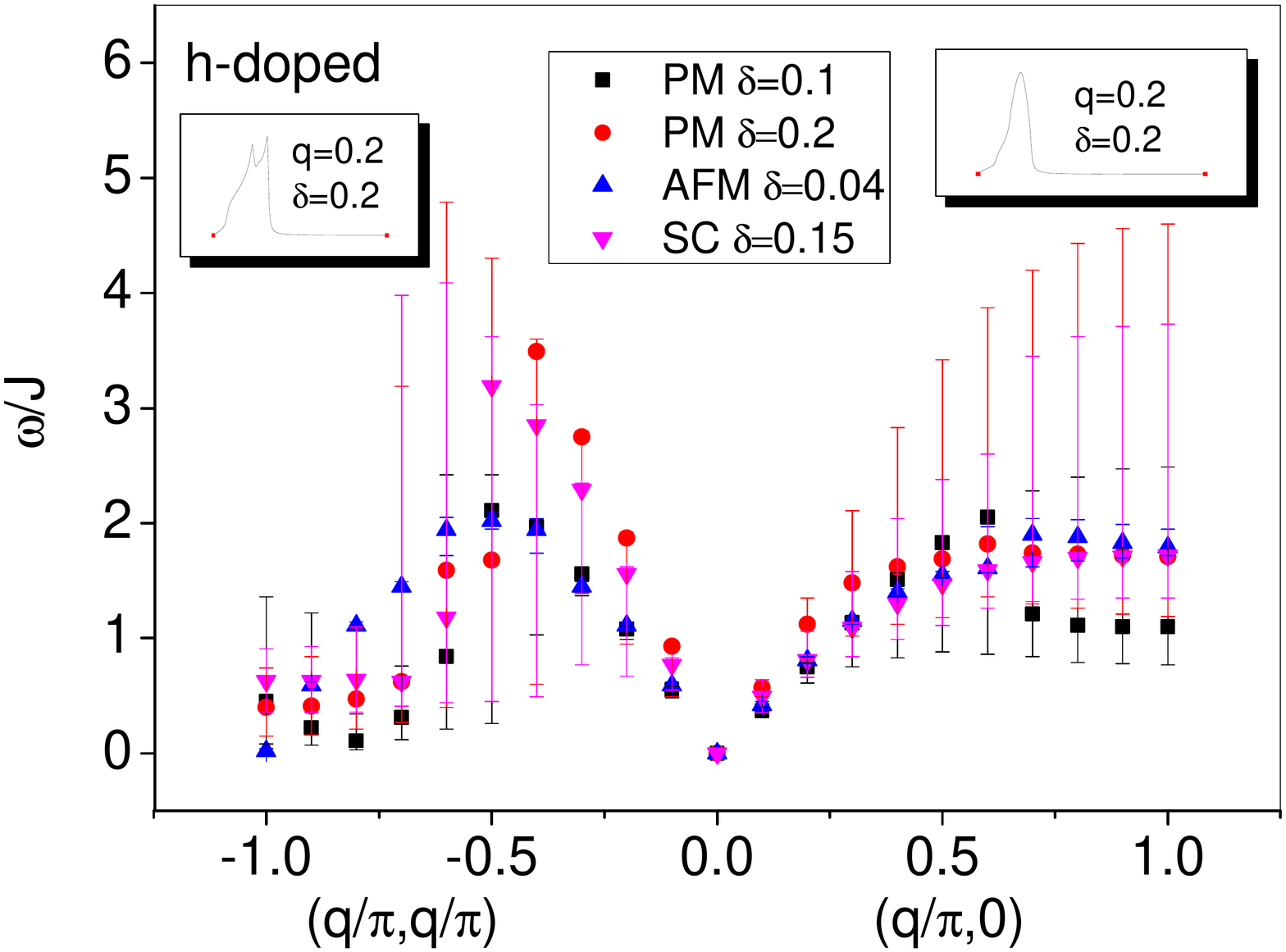}
\label{d-h} }
\subfigure[~dispersion for electron-doped cases]{
\includegraphics[width=.9\columnwidth]{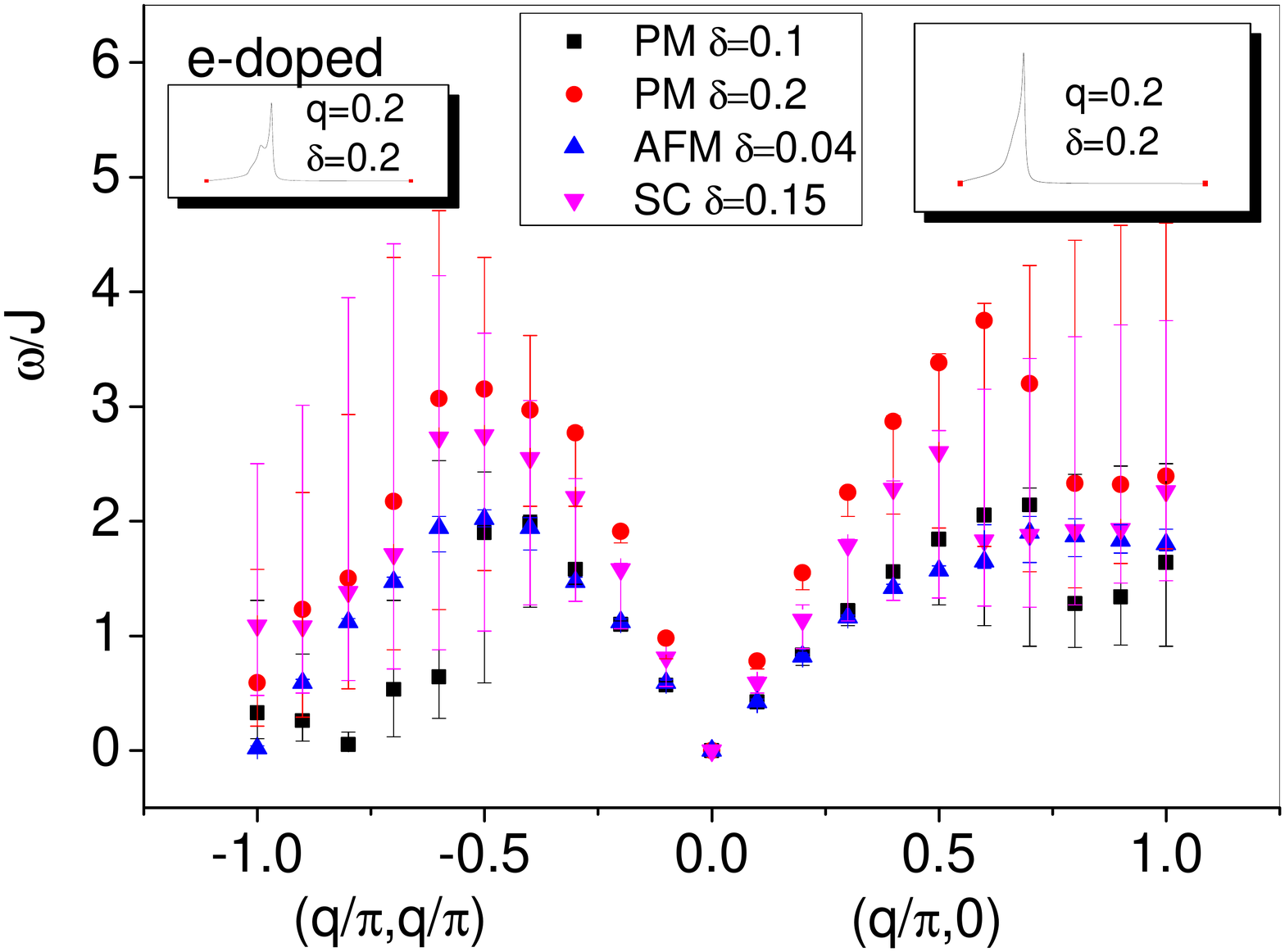}
\label{d-e} }
\caption{\label{Dispersion}The dispersion relation of the spin excitations in AFM and PM, (a) of the hole-doped case; (b) of the electron-doped case. The ends of the error bars indicate the half-maximum points. The points at the zero momentum are artificial in order to visualize the linearity. The electron-doped systems show the hardening effects along both directions while the hardening effect in the hole-doped systems along the $(\pi,0,)$ direction is reduced. Insets are the spin susceptibilities with (electron- or hole-) doping=0.2 and momentum transfer $q$=0.2 in both directions in k-space.}.
\end{figure}

For every momentum $q$ in the spin susceptibility, we identify the maximum of $\text{Im} \chi_{\text{RPA}}^\pm$ as the excitation energy. We plot the excitation energies with different momenta, thus the dispersion relation of the spin excitation, in Fig. \ref{Dispersion} for different hole concentrations in (a) and electrons in (b). We also indicate the half maxima of the susceptibility by the error bars.  In the high momentum region near q=($\pi,0$) and ($\pi/2,\pi/2$), the broadness of the spectrum makes the identification of the excitation energy difficult and causes large fluctuations as well as the feature of particle-hole excitations mentioned above.

In AFM cases, the dispersions are the collective spin-wave mode excitations, which agrees with experiments.\cite{Le-Tacon:2011fk,LeTacon2013,Dean2013} The dispersion at small q does not change significantly with the dopant density in the AFM phase. In Fig.\ref{Dispersion}(b), both AFM and PM results for the same electron concentration $\delta=0.1$ show very similar results at small $q$ and differ more significantly at larger $q$. But still the dispersions are very similar for both cases, showing that PM results at low $q$ are spin-wave-like excitations in nature.

In both electron-doped and hole-doped cases, the dispersion relation is linear for small $q$. The slopes of excitations at small $q$ increases with doping significantly for electron doped cases as shown in Fig. \ref{Dispersion}(b). This hardening effect was observed in recent experiments\cite{Lee2014}. Along the $q_x$ direction for the hole-doped cases, the slope is only slightly dependent on the dopant density (See the right panel of Fig. \ref{Dispersion}(a)), which is also qualitatively consistent with experiments\cite{Lee2014,Ishii2014,Le-Tacon:2011fk,LeTacon2013,Dean2013,Guarise2014,Dean2014,Dean2013b}. Insets in Fig. \ref{Dispersion} are the spin susceptibilities with (electron- or hole-) dopant density $\delta=0.2$ and momentum transfer $q$=0.2 in both directions in k-space. Only the hole-doped case along the $(\pi,\pi)$ direction shows two-peak feature, whose possible consequences will be discussed in section \ref{hole-pipi}. These doping dependences are due to  the doping-dependent bandwidth originated from the strong electron-electron correlations. This will be discussed in the following. 

Consider $\epsilon_k$ in Eq. \ref{MFH} along the $q_x$ direction. It is easy to see that the slope $s$ of the energy dispersion in the long wavelength region is proportional to the bandwidth. The bandwidth $W=4\delta(t+t')+2J\chi$ and $\frac{\Delta s}{\Delta \delta}\sim\frac{dW}{d\delta}=4(t+t')$. For the electron-doped cases, $t$ and $t'$ are of the same sign while for the hole-doped cases, $t$ and $t'$ have opposite signs. Thus the  bandwidth has a much larger dependence on dopant density for the electron-doped  cases than in the hole-doped cases. Similar analysis can be performed along the ($\pi,\pi$) direction. When the superconductivity exists, in addition to the particle-hole excitations, we also need to consider the particle-particle excitations. Estimating $\Delta_0$ as 0.2, the superconducting gap $\Delta_k\approx\frac{3J}{2}\Delta_0\approx0.3J$, which is small compared to the original band. Therefore, the excitation spectrum are not much altered in the presence of superconducting gap (See Fig. \ref{Dispersion}), which is consistent with recent experimental observation showing similar excitation dispersions at $T>T_c$\cite{Peng2015}.

\begin{figure}
   \includegraphics[scale =0.3] {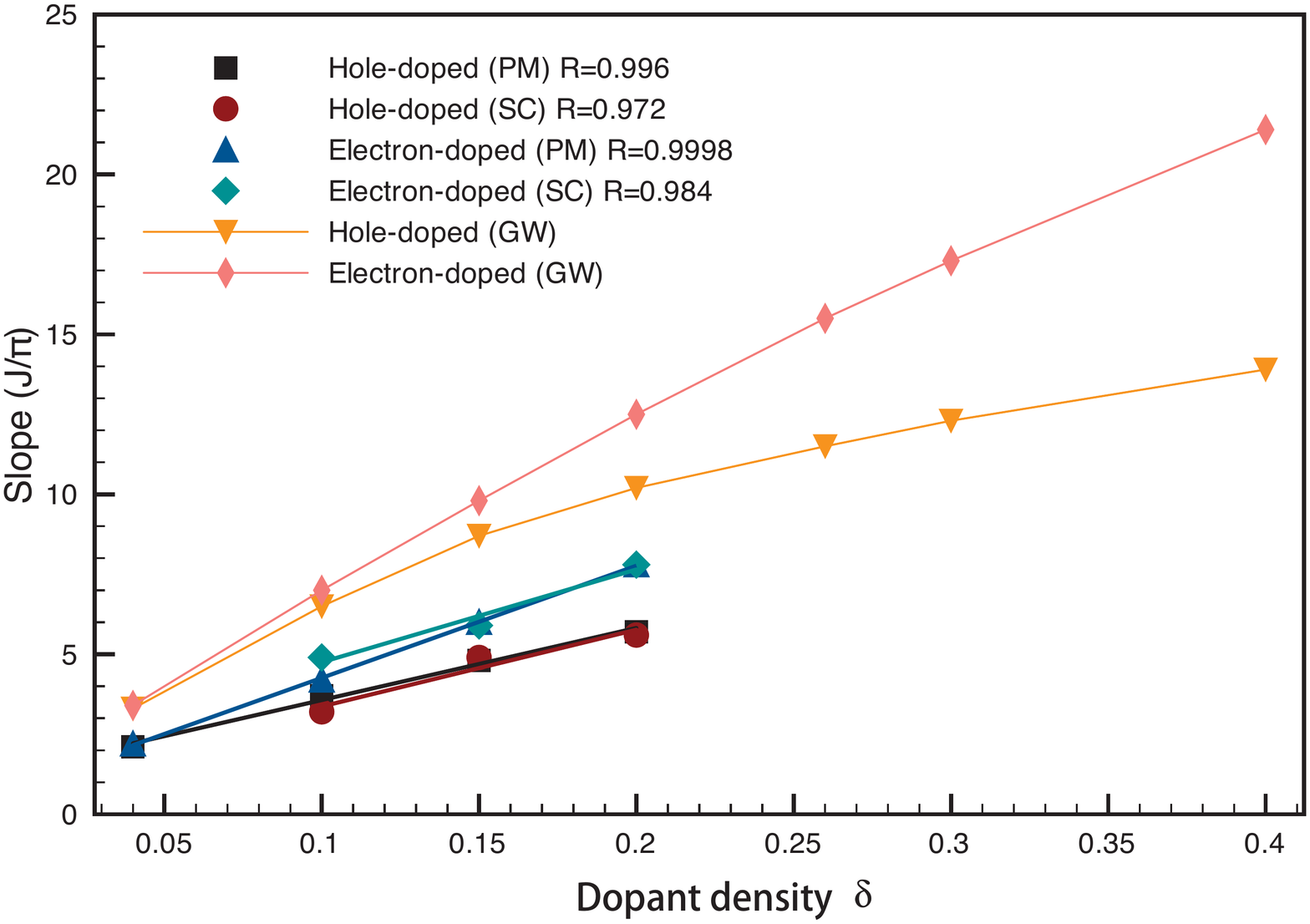}
   \caption{\label{slope} The slope near zero momentum of the dispersions as a functions of dopant density. The relations are almost linear, with the electron-doped cases ascends faster than the hole-doped cases as the doping increases. The case of Gutzwiller approximation (GW) is also plotted.}
\end{figure}

\begin{figure}
\includegraphics[width=.9\columnwidth]{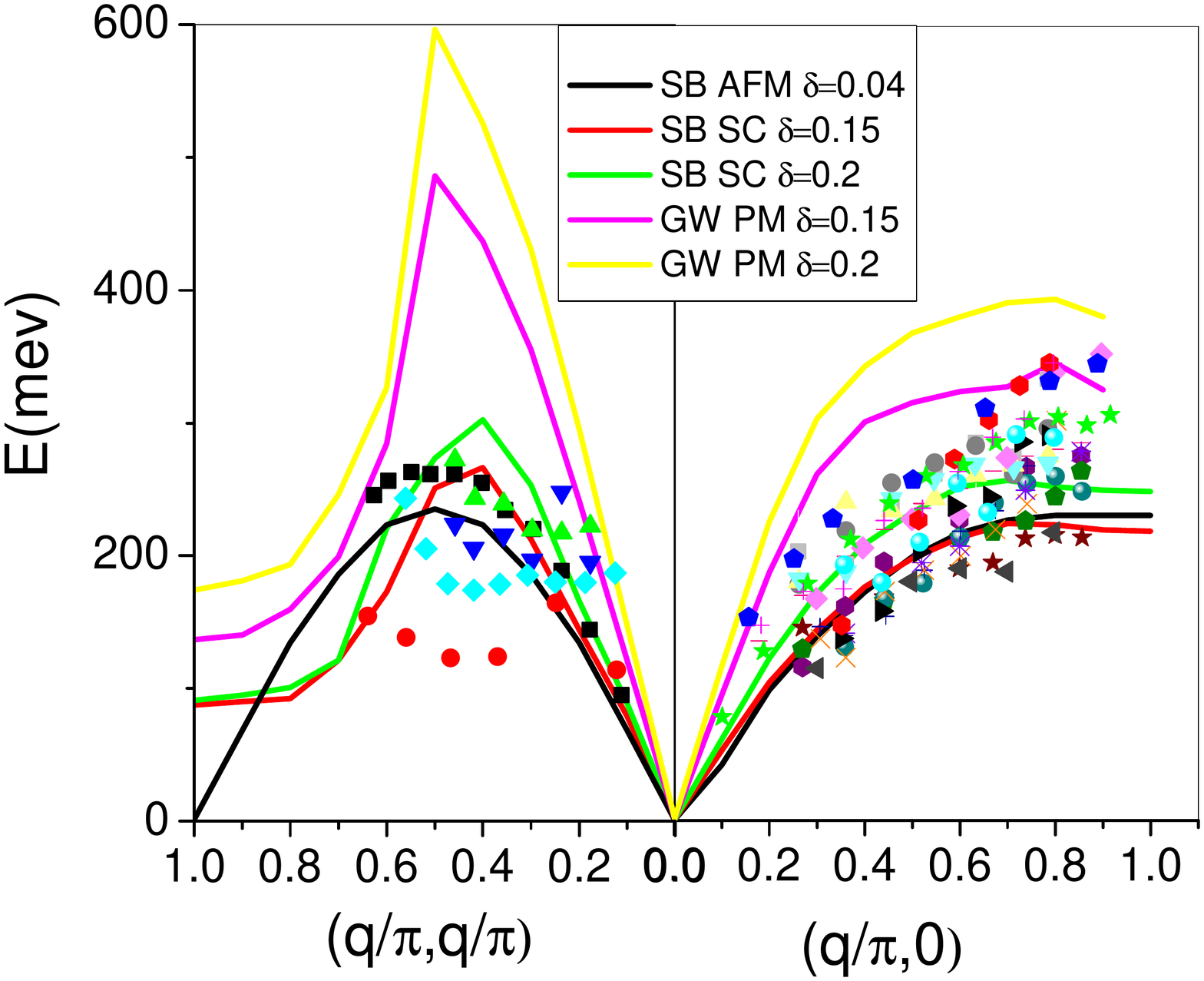}
\caption{\label{Exp} Comparison of excitations of our calculations using $J=120$ meV with experiments of hole-doped systems\cite{Le-Tacon:2011fk,LeTacon2013,Dean2013,Guarise2014,Dean2014,Dean2013b,Wakimoto2015,Huang2016} and HMHW$=50 meV$. SB stands for Slave-Boson method and GW stands for Gutzwiller.}
\end{figure}

In order to illustrate the effect of the renormalized bandwidth explicitly, we plot the dopant dependent slope of energy dispersion in the small momentum regime along the $(\pi,0)$ direction in Fig. \ref{slope}. We also include the Gutzwiller approximation (GW) by multiplying the hopping integrals by $g_t=2\delta/(1+\delta)$. Since GW factor is proportional to $2 \delta$  instead of just $\delta$ as in the Slave-Boson result, the slopes shown in Fig. \ref{slope} are about twice larger than that of Slave-Boson for both hole- and electron-doped cases. This confirms that the hardening is due to the band renormalization by the strong correlation. Since experiments for hole-doped systems have found similar dispersions\cite{Le-Tacon:2011fk,LeTacon2013,Dean2013,Guarise2014,Dean2014,Dean2013b} for  doping between 10\% to 40\% (See the right panel of Fig. \ref{Exp}), the reduced hardening in our hole-doped calculations indicates the strong correlation are still present for  doping as large as  40\%.

\begin{figure}.

\includegraphics[width=.9\columnwidth]{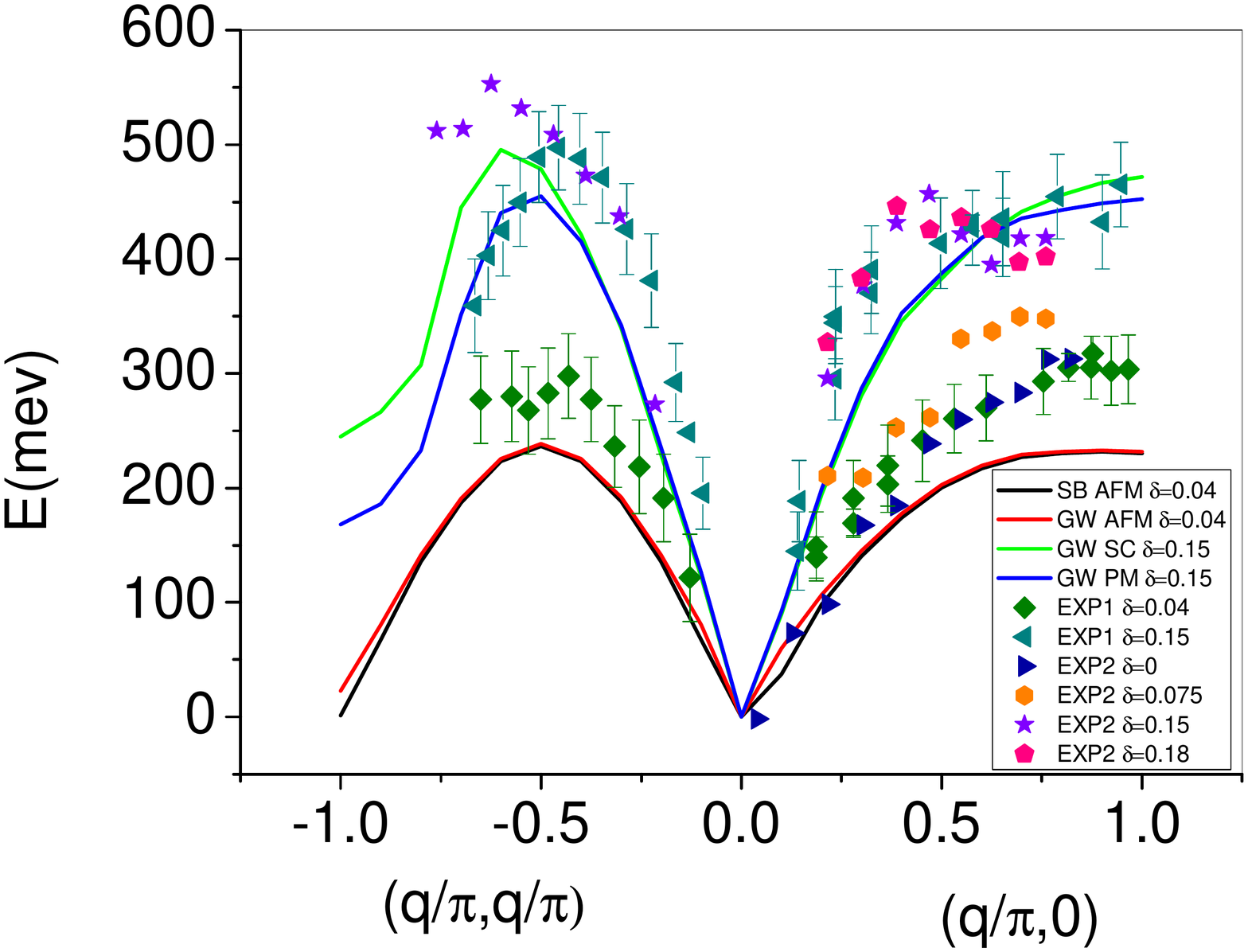}
\caption{\label{edopeexp} Comparison of energies of the energy excitations of our  calculations with the experiments (EXP1\cite{Lee2014} and EXP2\cite{Ishii2014}) of electron-doped NCCO with $J=120 meV$ and HMHW$=50 meV$. SB stands for Slave-Boson method and GW stands for Gutzwiller.}
\end{figure}

Our results are compared with experiments on the electron doped cuprates\cite{Lee2014,Ishii2014} in Fig.\ref{edopeexp}. In addition to the Slave-Boson method (SB), calculations with $\delta$ replaced by the Gutzwiller (GW) factor $2\delta/(1+\delta)$ are also carried out. The enhanced hardening effect in the experimental data  is well reproduced by  both Slave-Boson method (in AFM cases) and GW approximations (all dopings). This consistency with experimental observations is quite surprising as we have not included the core hole effect\cite{Ament2011}. Also, this provides a physical insight on the hardening effect before including the three-site term in the t-J model\cite{Jia2014}. Note that in AFM cases both methods give similar results even though the bandwidth in GW method is nearly twice of SB method. This is expected as the spin-wave excitations should dominate in the AFM regime.

In Fig.~\ref{Exp}, we compare our results with hole-doped experiments along the $(\pi,0)$ direction\cite{Le-Tacon:2011fk,LeTacon2013,Dean2013,Guarise2014,Dean2014,Dean2013b,Wakimoto2015,Huang2016} and along the $(\pi,\pi)$ direction\cite{Guarise2014,Wakimoto2015,Huang2016}. For the $(\pi,0)$ direction, our results are consistent with peaks and lineshapes reported by experiments and also having a similar energy spread. Along the $(\pi,\pi)$ direction, our dispersion using the GW factors is less consistent with experiments. One reason for this disagreement may be due to the large energy width in our calculations and most experimental data($\sim 300 meV$). Another possible reason is that it is difficult to determine the peak position as there are two peaks as shown in the left inset in Fig. \ref{Dispersion}. This issue  will be discussed more in the next section. All of our results are shown after applying the Gaussian convolution (See appendix \ref{GC}), the half-width at half-maximum of the distribution is $50$meV shown in Figs. \ref{edopeexp} and \ref{Exp}

\section{Discussions on hole-doped systems}
\label{hole-pipi}
Ref. \onlinecite{Jia2014} pointed out that while the hardening of the e-doped systems can be well explained by t-J like models, the hole-doped cases are not well-fitted and need the considerations of the full Hubbard model. The situation is similar with our calculation. While the reduced hardening effect along the $(\pi,0)$ direction in the hole-doped calculations are qualitatively consistent with the excitation spectrum in the experiments, the remarkable hardening effect along the $(\pi,\pi)$ direction is different from the experimental observations\cite{Guarise2014,Wakimoto2015,Huang2016}, indicating some key ingredients missed for this case in the simple SB+RPA calculations. 

In order to get more insights on this issue, we study the lineshape of the spin susceptibility in detail. We find that most lineshapes are of a well-defined one-peak structure while the two-peak structure appears in hole-doped systems along $(\pi,\pi)$ direction, as shown in insets in Fig. \ref{Dispersion}. Because these two peak values are close, upon introducing extra interactions or considering other possible effects, the larger peak of the two, defined as the excitation peak, may switch while those susceptibilities with one-peak structure may be relatively robust. This may change our calculated excitation dispersions. 

As an example, we consider an frequency-dependent lifetime, $\tau(\omega)$ of quasiparticles ($1/\tau\sim a+b\omega$ in the marginal-Fermi-liquid theory\cite{Varma1989} and $1/\tau\sim c+d\omega^2$ in the normal-Fermi-liquid theory.) in the mean-field SB stage. The inclusion of this variable lifetime switches some of the maximum of those spin susceptibilities with two-peak structures and leads to the nearly doping-independent excitation spectrum in the hole-doped cases along the $(\pi,\pi)$ direction (doping $=0.15\sim0.2$) while other cases (electron-doped systems and hole-doped systems along the $(\pi,0)$ direction) are almost unchanged by this inclusion, which fits better to experiments. 

Although only partially consistent with experiments, our theory does show the uniqueness of the hole-doped systems along the $(\pi,\pi)$ direction. We argue that by including some minor interactions or effects, which is out of the scope of our simple SB+RPA scheme here, the excitation dispersion for hole-doped systems along the $(\pi,\pi)$ direction can be modified and reach better consistency with experimental observations.

\section{Conclusions}
In summary, we investigate the spin-spin susceptibility in the $t-t'-t''-J$ model, via Slave-Boson mean-field theory. The excitation spectrum are determined through the peaks of the imaginary part of the susceptibility. The paramagnon hardening effect, consistent with experimental observations in electron-doped cuprates, comes from the doping dependent bandwidth, revealing the strong correlation. Nevertheless, this effect is lessened in the hole-doped materials, partly reflecting the nearly doping-independent energy dispersion in hole-doped experiments. We argue that discrepancies in the hole-doped systems along the $(\pi,\pi)$ direction may be fixed by including some minor interactions or effects. We also show that both particle-hole like  and the collective spin-wave like excitations are usually coupled together and not easily separated. The increase of bandwidth with dopant density due to the strong correlation is still present over a wide doping range in cuprates.

\acknowledgments
The authors acknowledge useful discussions with Sung-Po Chao (Academia Sinica, Taiwan). WJL, CJL and TKL acknowledge the support by Taiwan's MOST (MOST 104-2112-M-001-005).\\

\appendix
\section{Gaussian convolution}
\label{GC}
Due to the limited energy resolution in some of the RIXS experiments\cite{Dean2015}, we have to apply the Gaussian convolution to our calculation results before comparing them to the experimental observations. Gaussian function is defined as
\begin{equation}
 G(\omega)=\exp(-\frac{\omega^2}{2\sigma}).
\end{equation}
For every frequency $\omega$, the newly convoluted data are calculated by
\begin{equation}
\text{New Data}(\omega) = A\frac{\sum_n G(\omega-\omega_n)\times \text{Original Data}(\omega_n)}{\sum_n G(\omega-\omega_n)},
\end{equation}
where $n$ stands for summing all the frequency points, and $A$ is a normalization factor to keep the total weight conserved. As an example, suppose HWHM in the RIXS experiments is around $50\ meV$. The half width at half maximum (HWHM) is given by $\sqrt{2\ln 2} \sigma$ so that we set $\sigma=0.354J$ with $J=120\ meV$ in the Gaussian convolutions.

\bibliographystyle{apsrev4-1}
\bibliography{SpinExcitation}

\begin{thebibliography}{34}%
\makeatletter
\providecommand \@ifxundefined [1]{%
 \@ifx{#1\undefined}
}%
\providecommand \@ifnum [1]{%
 \ifnum #1\expandafter \@firstoftwo
 \else \expandafter \@secondoftwo
 \fi
}%
\providecommand \@ifx [1]{%
 \ifx #1\expandafter \@firstoftwo
 \else \expandafter \@secondoftwo
 \fi
}%
\providecommand \natexlab [1]{#1}%
\providecommand \enquote  [1]{``#1''}%
\providecommand \bibnamefont  [1]{#1}%
\providecommand \bibfnamefont [1]{#1}%
\providecommand \citenamefont [1]{#1}%
\providecommand \href@noop [0]{\@secondoftwo}%
\providecommand \href [0]{\begingroup \@sanitize@url \@href}%
\providecommand \@href[1]{\@@startlink{#1}\@@href}%
\providecommand \@@href[1]{\endgroup#1\@@endlink}%
\providecommand \@sanitize@url [0]{\catcode `\\12\catcode `\$12\catcode
  `\&12\catcode `\#12\catcode `\^12\catcode `\_12\catcode `\%12\relax}%
\providecommand \@@startlink[1]{}%
\providecommand \@@endlink[0]{}%
\providecommand \url  [0]{\begingroup\@sanitize@url \@url }%
\providecommand \@url [1]{\endgroup\@href {#1}{\urlprefix }}%
\providecommand \urlprefix  [0]{URL }%
\providecommand \Eprint [0]{\href }%
\providecommand \doibase [0]{http://dx.doi.org/}%
\providecommand \selectlanguage [0]{\@gobble}%
\providecommand \bibinfo  [0]{\@secondoftwo}%
\providecommand \bibfield  [0]{\@secondoftwo}%
\providecommand \translation [1]{[#1]}%
\providecommand \BibitemOpen [0]{}%
\providecommand \bibitemStop [0]{}%
\providecommand \bibitemNoStop [0]{.\EOS\space}%
\providecommand \EOS [0]{\spacefactor3000\relax}%
\providecommand \BibitemShut  [1]{\csname bibitem#1\endcsname}%
\let\auto@bib@innerbib\@empty
\bibitem [{\citenamefont {Lee}\ \emph {et~al.}(2006)\citenamefont {Lee},
  \citenamefont {Nagaosa},\ and\ \citenamefont {Wen}}]{Rev2006}%
  \BibitemOpen
  \bibfield  {author} {\bibinfo {author} {\bibfnamefont {P.~A.}\ \bibnamefont
  {Lee}}, \bibinfo {author} {\bibfnamefont {N.}~\bibnamefont {Nagaosa}}, \ and\
  \bibinfo {author} {\bibfnamefont {X.-G.}\ \bibnamefont {Wen}},\ }\href@noop
  {} {\bibfield  {journal} {\bibinfo  {journal} {Rev. Mod. Phys.}\ }\textbf
  {\bibinfo {volume} {78}},\ \bibinfo {pages} {17} (\bibinfo {year}
  {2006})}\BibitemShut {NoStop}%
\bibitem [{\citenamefont {Ament}\ \emph {et~al.}(2011)\citenamefont {Ament},
  \citenamefont {van Veenendaal}, \citenamefont {Devereaux}, \citenamefont
  {Hill},\ and\ \citenamefont {van~den Brink}}]{Ament2011}%
  \BibitemOpen
  \bibfield  {author} {\bibinfo {author} {\bibfnamefont {L.~J.~P.}\
  \bibnamefont {Ament}}, \bibinfo {author} {\bibfnamefont {M.}~\bibnamefont
  {van Veenendaal}}, \bibinfo {author} {\bibfnamefont {T.~P.}\ \bibnamefont
  {Devereaux}}, \bibinfo {author} {\bibfnamefont {J.~P.}\ \bibnamefont {Hill}},
  \ and\ \bibinfo {author} {\bibfnamefont {J.}~\bibnamefont {van~den Brink}},\
  }\href@noop {} {\bibfield  {journal} {\bibinfo  {journal} {Rev. Mod. Phys.}\
  }\textbf {\bibinfo {volume} {83}},\ \bibinfo {pages} {705} (\bibinfo {year}
  {2011})}\BibitemShut {NoStop}%
\bibitem [{\citenamefont {Dean}(2015)}]{Dean2015}%
  \BibitemOpen
  \bibfield  {author} {\bibinfo {author} {\bibfnamefont {M.}~\bibnamefont
  {Dean}},\ }\href@noop {} {\bibfield  {journal} {\bibinfo  {journal} {J. Magn.
  Magn. Mater.}\ }\textbf {\bibinfo {volume} {376}},\ \bibinfo {pages} {3}
  (\bibinfo {year} {2015})}\BibitemShut {NoStop}%
\bibitem [{\citenamefont {Le~Tacon}\ \emph {et~al.}(2011)\citenamefont
  {Le~Tacon}, \citenamefont {Ghiringhelli}, \citenamefont {Chaloupka},
  \citenamefont {Sala}, \citenamefont {Hinkov}, \citenamefont {Haverkort},
  \citenamefont {Minola}, \citenamefont {Bakr}, \citenamefont {Zhou},
  \citenamefont {Blanco-Canosa}, \citenamefont {Monney}, \citenamefont {Song},
  \citenamefont {Sun}, \citenamefont {Lin}, \citenamefont {De~Luca},
  \citenamefont {Salluzzo}, \citenamefont {Khaliullin}, \citenamefont
  {Schmitt}, \citenamefont {Braicovich},\ and\ \citenamefont
  {Keimer}}]{Le-Tacon:2011fk}%
  \BibitemOpen
  \bibfield  {author} {\bibinfo {author} {\bibfnamefont {M.}~\bibnamefont
  {Le~Tacon}}, \bibinfo {author} {\bibfnamefont {G.}~\bibnamefont
  {Ghiringhelli}}, \bibinfo {author} {\bibfnamefont {J.}~\bibnamefont
  {Chaloupka}}, \bibinfo {author} {\bibfnamefont {M.~M.}\ \bibnamefont {Sala}},
  \bibinfo {author} {\bibfnamefont {V.}~\bibnamefont {Hinkov}}, \bibinfo
  {author} {\bibfnamefont {M.~W.}\ \bibnamefont {Haverkort}}, \bibinfo {author}
  {\bibfnamefont {M.}~\bibnamefont {Minola}}, \bibinfo {author} {\bibfnamefont
  {M.}~\bibnamefont {Bakr}}, \bibinfo {author} {\bibfnamefont {K.~J.}\
  \bibnamefont {Zhou}}, \bibinfo {author} {\bibfnamefont {S.}~\bibnamefont
  {Blanco-Canosa}}, \bibinfo {author} {\bibfnamefont {C.}~\bibnamefont
  {Monney}}, \bibinfo {author} {\bibfnamefont {Y.~T.}\ \bibnamefont {Song}},
  \bibinfo {author} {\bibfnamefont {G.~L.}\ \bibnamefont {Sun}}, \bibinfo
  {author} {\bibfnamefont {C.~T.}\ \bibnamefont {Lin}}, \bibinfo {author}
  {\bibfnamefont {G.~M.}\ \bibnamefont {De~Luca}}, \bibinfo {author}
  {\bibfnamefont {M.}~\bibnamefont {Salluzzo}}, \bibinfo {author}
  {\bibfnamefont {G.}~\bibnamefont {Khaliullin}}, \bibinfo {author}
  {\bibfnamefont {T.}~\bibnamefont {Schmitt}}, \bibinfo {author} {\bibfnamefont
  {L.}~\bibnamefont {Braicovich}}, \ and\ \bibinfo {author} {\bibfnamefont
  {B.}~\bibnamefont {Keimer}},\ }\href@noop {} {\bibfield  {journal} {\bibinfo
  {journal} {Nat. Phys.}\ }\textbf {\bibinfo {volume} {7}},\ \bibinfo {pages}
  {725} (\bibinfo {year} {2011})}\BibitemShut {NoStop}%
\bibitem [{\citenamefont {Le~Tacon}\ \emph {et~al.}(2013)\citenamefont
  {Le~Tacon}, \citenamefont {Minola}, \citenamefont {Peets}, \citenamefont
  {Moretti~Sala}, \citenamefont {Blanco-Canosa}, \citenamefont {Hinkov},
  \citenamefont {Liang}, \citenamefont {Bonn}, \citenamefont {Hardy},
  \citenamefont {Lin}, \citenamefont {Schmitt}, \citenamefont {Braicovich},
  \citenamefont {Ghiringhelli},\ and\ \citenamefont {Keimer}}]{LeTacon2013}%
  \BibitemOpen
  \bibfield  {author} {\bibinfo {author} {\bibfnamefont {M.}~\bibnamefont
  {Le~Tacon}}, \bibinfo {author} {\bibfnamefont {M.}~\bibnamefont {Minola}},
  \bibinfo {author} {\bibfnamefont {D.~C.}\ \bibnamefont {Peets}}, \bibinfo
  {author} {\bibfnamefont {M.}~\bibnamefont {Moretti~Sala}}, \bibinfo {author}
  {\bibfnamefont {S.}~\bibnamefont {Blanco-Canosa}}, \bibinfo {author}
  {\bibfnamefont {V.}~\bibnamefont {Hinkov}}, \bibinfo {author} {\bibfnamefont
  {R.}~\bibnamefont {Liang}}, \bibinfo {author} {\bibfnamefont {D.~A.}\
  \bibnamefont {Bonn}}, \bibinfo {author} {\bibfnamefont {W.~N.}\ \bibnamefont
  {Hardy}}, \bibinfo {author} {\bibfnamefont {C.~T.}\ \bibnamefont {Lin}},
  \bibinfo {author} {\bibfnamefont {T.}~\bibnamefont {Schmitt}}, \bibinfo
  {author} {\bibfnamefont {L.}~\bibnamefont {Braicovich}}, \bibinfo {author}
  {\bibfnamefont {G.}~\bibnamefont {Ghiringhelli}}, \ and\ \bibinfo {author}
  {\bibfnamefont {B.}~\bibnamefont {Keimer}},\ }\href@noop {} {\bibfield
  {journal} {\bibinfo  {journal} {Phys. Rev. B}\ }\textbf {\bibinfo {volume}
  {88}},\ \bibinfo {pages} {020501} (\bibinfo {year} {2013})}\BibitemShut
  {NoStop}%
\bibitem [{\citenamefont {Dean}\ \emph
  {et~al.}(2013{\natexlab{a}})\citenamefont {Dean}, \citenamefont {Dellea},
  \citenamefont {Springell}, \citenamefont {Yakhou-Harris}, \citenamefont
  {Kummer}, \citenamefont {Brookes}, \citenamefont {Liu}, \citenamefont {Sun},
  \citenamefont {Strle}, \citenamefont {Schmitt}, \citenamefont {Braicovich},
  \citenamefont {Ghiringhelli}, \citenamefont {Bo\v{z}ovi\'{c}},\ and\
  \citenamefont {Hill}}]{Dean2013}%
  \BibitemOpen
  \bibfield  {author} {\bibinfo {author} {\bibfnamefont {M.~P.~M.}\
  \bibnamefont {Dean}}, \bibinfo {author} {\bibfnamefont {G.}~\bibnamefont
  {Dellea}}, \bibinfo {author} {\bibfnamefont {R.~S.}\ \bibnamefont
  {Springell}}, \bibinfo {author} {\bibfnamefont {F.}~\bibnamefont
  {Yakhou-Harris}}, \bibinfo {author} {\bibfnamefont {K.}~\bibnamefont
  {Kummer}}, \bibinfo {author} {\bibfnamefont {N.~B.}\ \bibnamefont {Brookes}},
  \bibinfo {author} {\bibfnamefont {X.}~\bibnamefont {Liu}}, \bibinfo {author}
  {\bibfnamefont {Y.-J.}\ \bibnamefont {Sun}}, \bibinfo {author} {\bibfnamefont
  {J.}~\bibnamefont {Strle}}, \bibinfo {author} {\bibfnamefont
  {T.}~\bibnamefont {Schmitt}}, \bibinfo {author} {\bibfnamefont
  {L.}~\bibnamefont {Braicovich}}, \bibinfo {author} {\bibfnamefont
  {G.}~\bibnamefont {Ghiringhelli}}, \bibinfo {author} {\bibfnamefont
  {I.}~\bibnamefont {Bo\v{z}ovi\'{c}}}, \ and\ \bibinfo {author} {\bibfnamefont
  {J.~P.}\ \bibnamefont {Hill}},\ }\href@noop {} {\bibfield  {journal}
  {\bibinfo  {journal} {Nat. Mater.}\ }\textbf {\bibinfo {volume} {12}},\
  \bibinfo {pages} {1019} (\bibinfo {year} {2013}{\natexlab{a}})}\BibitemShut
  {NoStop}%
\bibitem [{\citenamefont {Braicovich}\ \emph {et~al.}(2010)\citenamefont
  {Braicovich}, \citenamefont {van~den Brink}, \citenamefont {Bisogni},
  \citenamefont {Sala}, \citenamefont {Ament}, \citenamefont {Brookes},
  \citenamefont {De~Luca}, \citenamefont {Salluzzo}, \citenamefont {Schmitt},
  \citenamefont {Strocov},\ and\ \citenamefont
  {Ghiringhelli}}]{Braicovich2010}%
  \BibitemOpen
  \bibfield  {author} {\bibinfo {author} {\bibfnamefont {L.}~\bibnamefont
  {Braicovich}}, \bibinfo {author} {\bibfnamefont {J.}~\bibnamefont {van~den
  Brink}}, \bibinfo {author} {\bibfnamefont {V.}~\bibnamefont {Bisogni}},
  \bibinfo {author} {\bibfnamefont {M.~M.}\ \bibnamefont {Sala}}, \bibinfo
  {author} {\bibfnamefont {L.~J.~P.}\ \bibnamefont {Ament}}, \bibinfo {author}
  {\bibfnamefont {N.~B.}\ \bibnamefont {Brookes}}, \bibinfo {author}
  {\bibfnamefont {G.~M.}\ \bibnamefont {De~Luca}}, \bibinfo {author}
  {\bibfnamefont {M.}~\bibnamefont {Salluzzo}}, \bibinfo {author}
  {\bibfnamefont {T.}~\bibnamefont {Schmitt}}, \bibinfo {author} {\bibfnamefont
  {V.~N.}\ \bibnamefont {Strocov}}, \ and\ \bibinfo {author} {\bibfnamefont
  {G.}~\bibnamefont {Ghiringhelli}},\ }\href@noop {} {\bibfield  {journal}
  {\bibinfo  {journal} {Phys. Rev. Lett.}\ }\textbf {\bibinfo {volume} {104}},\
  \bibinfo {pages} {077002} (\bibinfo {year} {2010})}\BibitemShut {NoStop}%
\bibitem [{\citenamefont {Lee}\ \emph {et~al.}(2014)\citenamefont {Lee},
  \citenamefont {Lee}, \citenamefont {Nowadnick}, \citenamefont {Gerber},
  \citenamefont {Tabis}, \citenamefont {Huang}, \citenamefont {Strocov},
  \citenamefont {Motoyama}, \citenamefont {Yu}, \citenamefont {Moritz},
  \citenamefont {Huang}, \citenamefont {Wang}, \citenamefont {Huang},
  \citenamefont {Wu}, \citenamefont {Chen}, \citenamefont {Huang},
  \citenamefont {Greven}, \citenamefont {Schmitt}, \citenamefont {Shen},\ and\
  \citenamefont {Devereaux}}]{Lee2014}%
  \BibitemOpen
  \bibfield  {author} {\bibinfo {author} {\bibfnamefont {W.~S.}\ \bibnamefont
  {Lee}}, \bibinfo {author} {\bibfnamefont {J.}~\bibnamefont {Lee}}, \bibinfo
  {author} {\bibfnamefont {E.}~\bibnamefont {Nowadnick}}, \bibinfo {author}
  {\bibfnamefont {S.}~\bibnamefont {Gerber}}, \bibinfo {author} {\bibfnamefont
  {W.}~\bibnamefont {Tabis}}, \bibinfo {author} {\bibfnamefont
  {S.}~\bibnamefont {Huang}}, \bibinfo {author} {\bibfnamefont {V.~N.}\
  \bibnamefont {Strocov}}, \bibinfo {author} {\bibfnamefont {E.~M.}\
  \bibnamefont {Motoyama}}, \bibinfo {author} {\bibfnamefont {G.}~\bibnamefont
  {Yu}}, \bibinfo {author} {\bibfnamefont {B.}~\bibnamefont {Moritz}}, \bibinfo
  {author} {\bibfnamefont {H.}~\bibnamefont {Huang}}, \bibinfo {author}
  {\bibfnamefont {R.}~\bibnamefont {Wang}}, \bibinfo {author} {\bibfnamefont
  {Y.}~\bibnamefont {Huang}}, \bibinfo {author} {\bibfnamefont
  {W.}~\bibnamefont {Wu}}, \bibinfo {author} {\bibfnamefont {C.}~\bibnamefont
  {Chen}}, \bibinfo {author} {\bibfnamefont {D.}~\bibnamefont {Huang}},
  \bibinfo {author} {\bibfnamefont {M.}~\bibnamefont {Greven}}, \bibinfo
  {author} {\bibfnamefont {T.}~\bibnamefont {Schmitt}}, \bibinfo {author}
  {\bibfnamefont {Z.~X.}\ \bibnamefont {Shen}}, \ and\ \bibinfo {author}
  {\bibfnamefont {T.~P.}\ \bibnamefont {Devereaux}},\ }\href@noop {} {\bibfield
   {journal} {\bibinfo  {journal} {Nat. Phys.}\ }\textbf {\bibinfo {volume}
  {10}},\ \bibinfo {pages} {883} (\bibinfo {year} {2014})}\BibitemShut
  {NoStop}%
\bibitem [{\citenamefont {Wakimoto}\ \emph {et~al.}(2015)\citenamefont
  {Wakimoto}, \citenamefont {Ishii}, \citenamefont {Kimura}, \citenamefont
  {Fujita}, \citenamefont {Dellea}, \citenamefont {Kummer}, \citenamefont
  {Braicovich}, \citenamefont {Ghiringhelli}, \citenamefont {Debeer-Schmitt},\
  and\ \citenamefont {Granroth}}]{Wakimoto2015}%
  \BibitemOpen
  \bibfield  {author} {\bibinfo {author} {\bibfnamefont {S.}~\bibnamefont
  {Wakimoto}}, \bibinfo {author} {\bibfnamefont {K.}~\bibnamefont {Ishii}},
  \bibinfo {author} {\bibfnamefont {H.}~\bibnamefont {Kimura}}, \bibinfo
  {author} {\bibfnamefont {M.}~\bibnamefont {Fujita}}, \bibinfo {author}
  {\bibfnamefont {G.}~\bibnamefont {Dellea}}, \bibinfo {author} {\bibfnamefont
  {K.}~\bibnamefont {Kummer}}, \bibinfo {author} {\bibfnamefont
  {L.}~\bibnamefont {Braicovich}}, \bibinfo {author} {\bibfnamefont
  {G.}~\bibnamefont {Ghiringhelli}}, \bibinfo {author} {\bibfnamefont {L.~M.}\
  \bibnamefont {Debeer-Schmitt}}, \ and\ \bibinfo {author} {\bibfnamefont
  {G.~E.}\ \bibnamefont {Granroth}},\ }\href@noop {} {\bibfield  {journal}
  {\bibinfo  {journal} {Phys. Rev. B.}\ }\textbf {\bibinfo {volume} {91}},\
  \bibinfo {pages} {184513} (\bibinfo {year} {2015})}\BibitemShut {NoStop}%
\bibitem [{\citenamefont {Minola}\ \emph {et~al.}(2015)\citenamefont {Minola},
  \citenamefont {Dellea}, \citenamefont {Gretarsson}, \citenamefont {Peng},
  \citenamefont {Lu}, \citenamefont {Porras}, \citenamefont {Loew},
  \citenamefont {Yakhou}, \citenamefont {Brookes}, \citenamefont {Huang},
  \citenamefont {Pelliciari}, \citenamefont {Schmitt}, \citenamefont
  {Ghiringhelli}, \citenamefont {Keimer}, \citenamefont {Braicovich},\ and\
  \citenamefont {{Le Tacon}}}]{Minola2015}%
  \BibitemOpen
  \bibfield  {author} {\bibinfo {author} {\bibfnamefont {M.}~\bibnamefont
  {Minola}}, \bibinfo {author} {\bibfnamefont {G.}~\bibnamefont {Dellea}},
  \bibinfo {author} {\bibfnamefont {H.}~\bibnamefont {Gretarsson}}, \bibinfo
  {author} {\bibfnamefont {Y.}~\bibnamefont {Peng}}, \bibinfo {author}
  {\bibfnamefont {Y.}~\bibnamefont {Lu}}, \bibinfo {author} {\bibfnamefont
  {J.}~\bibnamefont {Porras}}, \bibinfo {author} {\bibfnamefont
  {T.}~\bibnamefont {Loew}}, \bibinfo {author} {\bibfnamefont {F.}~\bibnamefont
  {Yakhou}}, \bibinfo {author} {\bibfnamefont {N.}~\bibnamefont {Brookes}},
  \bibinfo {author} {\bibfnamefont {Y.}~\bibnamefont {Huang}}, \bibinfo
  {author} {\bibfnamefont {J.}~\bibnamefont {Pelliciari}}, \bibinfo {author}
  {\bibfnamefont {T.}~\bibnamefont {Schmitt}}, \bibinfo {author} {\bibfnamefont
  {G.}~\bibnamefont {Ghiringhelli}}, \bibinfo {author} {\bibfnamefont
  {B.}~\bibnamefont {Keimer}}, \bibinfo {author} {\bibfnamefont
  {L.}~\bibnamefont {Braicovich}}, \ and\ \bibinfo {author} {\bibfnamefont
  {M.}~\bibnamefont {{Le Tacon}}},\ }\href@noop {} {\bibfield  {journal}
  {\bibinfo  {journal} {Phys. Rev. Lett.}\ }\textbf {\bibinfo {volume} {114}},\
  \bibinfo {pages} {217003} (\bibinfo {year} {2015})}\BibitemShut {NoStop}%
\bibitem [{\citenamefont {Guarise}\ \emph {et~al.}(2014)\citenamefont
  {Guarise}, \citenamefont {Piazza}, \citenamefont {Berger}, \citenamefont
  {Giannini}, \citenamefont {Schmitt}, \citenamefont {R\o~nnow}, \citenamefont
  {Sawatzky}, \citenamefont {van~den Brink}, \citenamefont {Altenfeld},
  \citenamefont {Eremin},\ and\ \citenamefont {Grioni}}]{Guarise2014}%
  \BibitemOpen
  \bibfield  {author} {\bibinfo {author} {\bibfnamefont {M.}~\bibnamefont
  {Guarise}}, \bibinfo {author} {\bibfnamefont {B.~D.}\ \bibnamefont {Piazza}},
  \bibinfo {author} {\bibfnamefont {H.}~\bibnamefont {Berger}}, \bibinfo
  {author} {\bibfnamefont {E.}~\bibnamefont {Giannini}}, \bibinfo {author}
  {\bibfnamefont {T.}~\bibnamefont {Schmitt}}, \bibinfo {author} {\bibfnamefont
  {H.~M.}\ \bibnamefont {R\o~nnow}}, \bibinfo {author} {\bibfnamefont {G.~a.}\
  \bibnamefont {Sawatzky}}, \bibinfo {author} {\bibfnamefont {J.}~\bibnamefont
  {van~den Brink}}, \bibinfo {author} {\bibfnamefont {D.}~\bibnamefont
  {Altenfeld}}, \bibinfo {author} {\bibfnamefont {I.}~\bibnamefont {Eremin}}, \
  and\ \bibinfo {author} {\bibfnamefont {M.}~\bibnamefont {Grioni}},\
  }\href@noop {} {\bibfield  {journal} {\bibinfo  {journal} {Nat. Commun.}\
  }\textbf {\bibinfo {volume} {5}},\ \bibinfo {pages} {5760} (\bibinfo {year}
  {2014})}\BibitemShut {NoStop}%
\bibitem [{\citenamefont {Dean}\ \emph {et~al.}(2014)\citenamefont {Dean},
  \citenamefont {James}, \citenamefont {Walters}, \citenamefont {Bisogni},
  \citenamefont {Jarrige}, \citenamefont {H\"{u}cker}, \citenamefont
  {Giannini}, \citenamefont {Fujita}, \citenamefont {Pelliciari}, \citenamefont
  {Huang}, \citenamefont {Konik}, \citenamefont {Schmitt},\ and\ \citenamefont
  {Hill}}]{Dean2014}%
  \BibitemOpen
  \bibfield  {author} {\bibinfo {author} {\bibfnamefont {M.~P.~M.}\
  \bibnamefont {Dean}}, \bibinfo {author} {\bibfnamefont {a.~J.~a.}\
  \bibnamefont {James}}, \bibinfo {author} {\bibfnamefont {a.~C.}\ \bibnamefont
  {Walters}}, \bibinfo {author} {\bibfnamefont {V.}~\bibnamefont {Bisogni}},
  \bibinfo {author} {\bibfnamefont {I.}~\bibnamefont {Jarrige}}, \bibinfo
  {author} {\bibfnamefont {M.}~\bibnamefont {H\"{u}cker}}, \bibinfo {author}
  {\bibfnamefont {E.}~\bibnamefont {Giannini}}, \bibinfo {author}
  {\bibfnamefont {M.}~\bibnamefont {Fujita}}, \bibinfo {author} {\bibfnamefont
  {J.}~\bibnamefont {Pelliciari}}, \bibinfo {author} {\bibfnamefont {Y.~B.}\
  \bibnamefont {Huang}}, \bibinfo {author} {\bibfnamefont {R.~M.}\ \bibnamefont
  {Konik}}, \bibinfo {author} {\bibfnamefont {T.}~\bibnamefont {Schmitt}}, \
  and\ \bibinfo {author} {\bibfnamefont {J.~P.}\ \bibnamefont {Hill}},\
  }\href@noop {} {\bibfield  {journal} {\bibinfo  {journal} {Phys. Rev. B.}\
  }\textbf {\bibinfo {volume} {90}},\ \bibinfo {pages} {220506R} (\bibinfo
  {year} {2014})}\BibitemShut {NoStop}%
\bibitem [{\citenamefont {Dean}\ \emph
  {et~al.}(2013{\natexlab{b}})\citenamefont {Dean}, \citenamefont {James},
  \citenamefont {Springell}, \citenamefont {Liu}, \citenamefont {Monney},
  \citenamefont {Zhou}, \citenamefont {Konik}, \citenamefont {Wen},
  \citenamefont {Xu}, \citenamefont {Gu}, \citenamefont {Strocov},
  \citenamefont {Schmitt},\ and\ \citenamefont {Hill}}]{Dean2013b}%
  \BibitemOpen
  \bibfield  {author} {\bibinfo {author} {\bibfnamefont {M.}~\bibnamefont
  {Dean}}, \bibinfo {author} {\bibfnamefont {a.}~\bibnamefont {James}},
  \bibinfo {author} {\bibfnamefont {R.}~\bibnamefont {Springell}}, \bibinfo
  {author} {\bibfnamefont {X.}~\bibnamefont {Liu}}, \bibinfo {author}
  {\bibfnamefont {C.}~\bibnamefont {Monney}}, \bibinfo {author} {\bibfnamefont
  {K.}~\bibnamefont {Zhou}}, \bibinfo {author} {\bibfnamefont {R.}~\bibnamefont
  {Konik}}, \bibinfo {author} {\bibfnamefont {J.}~\bibnamefont {Wen}}, \bibinfo
  {author} {\bibfnamefont {Z.}~\bibnamefont {Xu}}, \bibinfo {author}
  {\bibfnamefont {G.}~\bibnamefont {Gu}}, \bibinfo {author} {\bibfnamefont
  {V.}~\bibnamefont {Strocov}}, \bibinfo {author} {\bibfnamefont
  {T.}~\bibnamefont {Schmitt}}, \ and\ \bibinfo {author} {\bibfnamefont
  {J.}~\bibnamefont {Hill}},\ }\href@noop {} {\bibfield  {journal} {\bibinfo
  {journal} {Phys. Rev. Lett.}\ }\textbf {\bibinfo {volume} {110}},\ \bibinfo
  {pages} {147001} (\bibinfo {year} {2013}{\natexlab{b}})}\BibitemShut
  {NoStop}%
\bibitem [{\citenamefont {Monney}\ \emph {et~al.}(2016)\citenamefont {Monney},
  \citenamefont {Schmitt}, \citenamefont {Matt}, \citenamefont {Mesot},
  \citenamefont {Strocov}, \citenamefont {Lipscombe}, \citenamefont {Hayden},\
  and\ \citenamefont {Chang}}]{Monney2016}%
  \BibitemOpen
  \bibfield  {author} {\bibinfo {author} {\bibfnamefont {C.}~\bibnamefont
  {Monney}}, \bibinfo {author} {\bibfnamefont {T.}~\bibnamefont {Schmitt}},
  \bibinfo {author} {\bibfnamefont {C.~E.}\ \bibnamefont {Matt}}, \bibinfo
  {author} {\bibfnamefont {J.}~\bibnamefont {Mesot}}, \bibinfo {author}
  {\bibfnamefont {V.~N.}\ \bibnamefont {Strocov}}, \bibinfo {author}
  {\bibfnamefont {O.~J.}\ \bibnamefont {Lipscombe}}, \bibinfo {author}
  {\bibfnamefont {S.~M.}\ \bibnamefont {Hayden}}, \ and\ \bibinfo {author}
  {\bibfnamefont {J.}~\bibnamefont {Chang}},\ }\href@noop {} {\bibfield
  {journal} {\bibinfo  {journal} {Phys. Rev. B}\ }\textbf {\bibinfo {volume}
  {93}},\ \bibinfo {pages} {075103} (\bibinfo {year} {2016})}\BibitemShut
  {NoStop}%
\bibitem [{\citenamefont {Huang}\ \emph {et~al.}(2016)\citenamefont {Huang},
  \citenamefont {Jia}, \citenamefont {Chen}, \citenamefont {Wohlfeld},
  \citenamefont {Moritz}, \citenamefont {Devereaux}, \citenamefont {Wu},
  \citenamefont {Okamoto}, \citenamefont {Lee}, \citenamefont {Hashimoto},
  \citenamefont {He}, \citenamefont {Shen}, \citenamefont {Yoshida},
  \citenamefont {Eisaki}, \citenamefont {Mou}, \citenamefont {Chen},\ and\
  \citenamefont {Huang}}]{Huang2016}%
  \BibitemOpen
  \bibfield  {author} {\bibinfo {author} {\bibfnamefont {H.~Y.}\ \bibnamefont
  {Huang}}, \bibinfo {author} {\bibfnamefont {C.~J.}\ \bibnamefont {Jia}},
  \bibinfo {author} {\bibfnamefont {Z.~Y.}\ \bibnamefont {Chen}}, \bibinfo
  {author} {\bibfnamefont {K.}~\bibnamefont {Wohlfeld}}, \bibinfo {author}
  {\bibfnamefont {B.}~\bibnamefont {Moritz}}, \bibinfo {author} {\bibfnamefont
  {T.~P.}\ \bibnamefont {Devereaux}}, \bibinfo {author} {\bibfnamefont {W.~B.}\
  \bibnamefont {Wu}}, \bibinfo {author} {\bibfnamefont {J.}~\bibnamefont
  {Okamoto}}, \bibinfo {author} {\bibfnamefont {W.~S.}\ \bibnamefont {Lee}},
  \bibinfo {author} {\bibfnamefont {M.}~\bibnamefont {Hashimoto}}, \bibinfo
  {author} {\bibfnamefont {Y.}~\bibnamefont {He}}, \bibinfo {author}
  {\bibfnamefont {Z.~X.}\ \bibnamefont {Shen}}, \bibinfo {author}
  {\bibfnamefont {Y.}~\bibnamefont {Yoshida}}, \bibinfo {author} {\bibfnamefont
  {H.}~\bibnamefont {Eisaki}}, \bibinfo {author} {\bibfnamefont {C.~Y.}\
  \bibnamefont {Mou}}, \bibinfo {author} {\bibfnamefont {C.~T.}\ \bibnamefont
  {Chen}}, \ and\ \bibinfo {author} {\bibfnamefont {D.~J.}\ \bibnamefont
  {Huang}},\ }\href@noop {} {\bibfield  {journal} {\bibinfo  {journal} {Sci.
  Rep.}\ }\textbf {\bibinfo {volume} {{6}}},\ \bibinfo {pages} {19657}
  (\bibinfo {year} {{2016}})}\BibitemShut {NoStop}%
\bibitem [{\citenamefont {Zeyher}\ and\ \citenamefont
  {Greco}(2013)}]{Roland2013}%
  \BibitemOpen
  \bibfield  {author} {\bibinfo {author} {\bibfnamefont {R.}~\bibnamefont
  {Zeyher}}\ and\ \bibinfo {author} {\bibfnamefont {A.}~\bibnamefont {Greco}},\
  }\href@noop {} {\bibfield  {journal} {\bibinfo  {journal} {Phys. Rev. B}\
  }\textbf {\bibinfo {volume} {87}},\ \bibinfo {pages} {224511} (\bibinfo
  {year} {2013})}\BibitemShut {NoStop}%
\bibitem [{\citenamefont {Si}\ \emph {et~al.}(2016)\citenamefont {Si},
  \citenamefont {Yu},\ and\ \citenamefont {Abrahams}}]{Si2016}%
  \BibitemOpen
  \bibfield  {author} {\bibinfo {author} {\bibfnamefont {Q.}~\bibnamefont
  {Si}}, \bibinfo {author} {\bibfnamefont {R.}~\bibnamefont {Yu}}, \ and\
  \bibinfo {author} {\bibfnamefont {E.}~\bibnamefont {Abrahams}},\ }\href@noop
  {} {\bibfield  {journal} {\bibinfo  {journal} {Nature Reviews Materials}\ ,\
  \bibinfo {pages} {16017}} (\bibinfo {year} {2016})}\BibitemShut {NoStop}%
\bibitem [{\citenamefont {Jia}\ \emph {et~al.}(2014)\citenamefont {Jia},
  \citenamefont {Nowadnick}, \citenamefont {Wohlfeld}, \citenamefont {Kung},
  \citenamefont {Chen}, \citenamefont {Johnston}, \citenamefont {Tohyama},
  \citenamefont {Moritz},\ and\ \citenamefont {Devereaux}}]{Jia2014}%
  \BibitemOpen
  \bibfield  {author} {\bibinfo {author} {\bibfnamefont {C.~J.}\ \bibnamefont
  {Jia}}, \bibinfo {author} {\bibfnamefont {E.~a.}\ \bibnamefont {Nowadnick}},
  \bibinfo {author} {\bibfnamefont {K.}~\bibnamefont {Wohlfeld}}, \bibinfo
  {author} {\bibfnamefont {Y.~F.}\ \bibnamefont {Kung}}, \bibinfo {author}
  {\bibfnamefont {C.-C.}\ \bibnamefont {Chen}}, \bibinfo {author}
  {\bibfnamefont {S.}~\bibnamefont {Johnston}}, \bibinfo {author}
  {\bibfnamefont {T.}~\bibnamefont {Tohyama}}, \bibinfo {author} {\bibfnamefont
  {B.}~\bibnamefont {Moritz}}, \ and\ \bibinfo {author} {\bibfnamefont {T.~P.}\
  \bibnamefont {Devereaux}},\ }\href@noop {} {\bibfield  {journal} {\bibinfo
  {journal} {Nat. Commun.}\ }\textbf {\bibinfo {volume} {5}},\ \bibinfo {pages}
  {3314} (\bibinfo {year} {2014})}\BibitemShut {NoStop}%
\bibitem [{\citenamefont {Zhang}\ \emph {et~al.}(1988)\citenamefont {Zhang},
  \citenamefont {Gros}, \citenamefont {Rice},\ and\ \citenamefont
  {Shiba}}]{Zhang:1988uq}%
  \BibitemOpen
  \bibfield  {author} {\bibinfo {author} {\bibfnamefont {F.}~\bibnamefont
  {Zhang}}, \bibinfo {author} {\bibfnamefont {C.}~\bibnamefont {Gros}},
  \bibinfo {author} {\bibfnamefont {T.}~\bibnamefont {Rice}}, \ and\ \bibinfo
  {author} {\bibfnamefont {H.}~\bibnamefont {Shiba}},\ }\href@noop {}
  {\bibfield  {journal} {\bibinfo  {journal} {Supercon. Sci. Tech.}\ }\textbf
  {\bibinfo {volume} {1}},\ \bibinfo {pages} {36} (\bibinfo {year}
  {1988})}\BibitemShut {NoStop}%
\bibitem [{\citenamefont {Lee}\ and\ \citenamefont {Nagaosa}(1992)}]{Lee1992}%
  \BibitemOpen
  \bibfield  {author} {\bibinfo {author} {\bibfnamefont {P.~A.}\ \bibnamefont
  {Lee}}\ and\ \bibinfo {author} {\bibfnamefont {N.}~\bibnamefont {Nagaosa}},\
  }\href@noop {} {\bibfield  {journal} {\bibinfo  {journal} {Phys. Rev. B}\
  }\textbf {\bibinfo {volume} {46}},\ \bibinfo {pages} {5621} (\bibinfo {year}
  {1992})}\BibitemShut {NoStop}%
\bibitem [{\citenamefont {Bickers}(1987)}]{Bickers1987}%
  \BibitemOpen
  \bibfield  {author} {\bibinfo {author} {\bibfnamefont {N.~E.}\ \bibnamefont
  {Bickers}},\ }\href@noop {} {\bibfield  {journal} {\bibinfo  {journal} {Rev.
  Mod. Phys.}\ }\textbf {\bibinfo {volume} {59}},\ \bibinfo {pages} {845}
  (\bibinfo {year} {1987})}\BibitemShut {NoStop}%
\bibitem [{\citenamefont {Yuan}\ \emph {et~al.}(2004)\citenamefont {Yuan},
  \citenamefont {Chen}, \citenamefont {Lee},\ and\ \citenamefont
  {Ting}}]{Yuan2004}%
  \BibitemOpen
  \bibfield  {author} {\bibinfo {author} {\bibfnamefont {Q.}~\bibnamefont
  {Yuan}}, \bibinfo {author} {\bibfnamefont {Y.}~\bibnamefont {Chen}}, \bibinfo
  {author} {\bibfnamefont {T.~K.}\ \bibnamefont {Lee}}, \ and\ \bibinfo
  {author} {\bibfnamefont {C.~S.}\ \bibnamefont {Ting}},\ }\href@noop {}
  {\bibfield  {journal} {\bibinfo  {journal} {Phys. Rev. B}\ }\textbf {\bibinfo
  {volume} {69}},\ \bibinfo {pages} {214523} (\bibinfo {year}
  {2004})}\BibitemShut {NoStop}%
\bibitem [{\citenamefont {Yuan}\ \emph {et~al.}(2005)\citenamefont {Yuan},
  \citenamefont {Lee},\ and\ \citenamefont {Ting}}]{Yuan2005}%
  \BibitemOpen
  \bibfield  {author} {\bibinfo {author} {\bibfnamefont {Q.}~\bibnamefont
  {Yuan}}, \bibinfo {author} {\bibfnamefont {T.~K.}\ \bibnamefont {Lee}}, \
  and\ \bibinfo {author} {\bibfnamefont {C.~S.}\ \bibnamefont {Ting}},\
  }\href@noop {} {\bibfield  {journal} {\bibinfo  {journal} {Phys. Rev. B}\
  }\textbf {\bibinfo {volume} {71}},\ \bibinfo {pages} {134522} (\bibinfo
  {year} {2005})}\BibitemShut {NoStop}%
\bibitem [{\citenamefont {Lee}\ \emph {et~al.}(2003)\citenamefont {Lee},
  \citenamefont {Ho},\ and\ \citenamefont {Nagaosa}}]{TK2003}%
  \BibitemOpen
  \bibfield  {author} {\bibinfo {author} {\bibfnamefont {T.~K.}\ \bibnamefont
  {Lee}}, \bibinfo {author} {\bibfnamefont {C.-M.}\ \bibnamefont {Ho}}, \ and\
  \bibinfo {author} {\bibfnamefont {N.}~\bibnamefont {Nagaosa}},\ }\href@noop
  {} {\bibfield  {journal} {\bibinfo  {journal} {Phys. Rev. Lett.}\ }\textbf
  {\bibinfo {volume} {90}},\ \bibinfo {pages} {067001} (\bibinfo {year}
  {2003})}\BibitemShut {NoStop}%
\bibitem [{\citenamefont {Li}\ \emph {et~al.}(2000)\citenamefont {Li},
  \citenamefont {Mou},\ and\ \citenamefont {Lee}}]{Li2000}%
  \BibitemOpen
  \bibfield  {author} {\bibinfo {author} {\bibfnamefont {J.-X.}\ \bibnamefont
  {Li}}, \bibinfo {author} {\bibfnamefont {C.-Y.}\ \bibnamefont {Mou}}, \ and\
  \bibinfo {author} {\bibfnamefont {T.~K.}\ \bibnamefont {Lee}},\ }\href@noop
  {} {\bibfield  {journal} {\bibinfo  {journal} {Phys. Rev. B}\ }\textbf
  {\bibinfo {volume} {62}},\ \bibinfo {pages} {640} (\bibinfo {year}
  {2000})}\BibitemShut {NoStop}%
\bibitem [{\citenamefont {Li}\ \emph {et~al.}(2001)\citenamefont {Li},
  \citenamefont {Mou}, \citenamefont {Gong},\ and\ \citenamefont
  {Lee}}]{Li2001}%
  \BibitemOpen
  \bibfield  {author} {\bibinfo {author} {\bibfnamefont {J.-X.}\ \bibnamefont
  {Li}}, \bibinfo {author} {\bibfnamefont {C.-Y.}\ \bibnamefont {Mou}},
  \bibinfo {author} {\bibfnamefont {C.-D.}\ \bibnamefont {Gong}}, \ and\
  \bibinfo {author} {\bibfnamefont {T.~K.}\ \bibnamefont {Lee}},\ }\href@noop
  {} {\bibfield  {journal} {\bibinfo  {journal} {Phys. Rev. B}\ }\textbf
  {\bibinfo {volume} {64}},\ \bibinfo {pages} {104518} (\bibinfo {year}
  {2001})}\BibitemShut {NoStop}%
\bibitem [{\citenamefont {Li}\ and\ \citenamefont {Gong}(2002)}]{Li2002}%
  \BibitemOpen
  \bibfield  {author} {\bibinfo {author} {\bibfnamefont {J.-X.}\ \bibnamefont
  {Li}}\ and\ \bibinfo {author} {\bibfnamefont {C.-D.}\ \bibnamefont {Gong}},\
  }\href@noop {} {\bibfield  {journal} {\bibinfo  {journal} {Phys. Rev. B}\
  }\textbf {\bibinfo {volume} {66}},\ \bibinfo {pages} {014506} (\bibinfo
  {year} {2002})}\BibitemShut {NoStop}%
\bibitem [{\citenamefont {Li}\ \emph {et~al.}(2003)\citenamefont {Li},
  \citenamefont {Zhang},\ and\ \citenamefont {Luo}}]{Li2003}%
  \BibitemOpen
  \bibfield  {author} {\bibinfo {author} {\bibfnamefont {J.-X.}\ \bibnamefont
  {Li}}, \bibinfo {author} {\bibfnamefont {J.}~\bibnamefont {Zhang}}, \ and\
  \bibinfo {author} {\bibfnamefont {J.}~\bibnamefont {Luo}},\ }\href@noop {}
  {\bibfield  {journal} {\bibinfo  {journal} {Phys. Rev. B}\ }\textbf {\bibinfo
  {volume} {68}},\ \bibinfo {pages} {224503} (\bibinfo {year}
  {2003})}\BibitemShut {NoStop}%
\bibitem [{\citenamefont {Ubbens}\ and\ \citenamefont
  {Lee}(1992)}]{Patrick1992}%
  \BibitemOpen
  \bibfield  {author} {\bibinfo {author} {\bibfnamefont {M.~U.}\ \bibnamefont
  {Ubbens}}\ and\ \bibinfo {author} {\bibfnamefont {P.~A.}\ \bibnamefont
  {Lee}},\ }\href@noop {} {\bibfield  {journal} {\bibinfo  {journal} {Phys.
  Rev. B}\ }\textbf {\bibinfo {volume} {46}},\ \bibinfo {pages} {8434}
  (\bibinfo {year} {1992})}\BibitemShut {NoStop}%
\bibitem [{\citenamefont {Brinckmann}\ and\ \citenamefont
  {Lee}(1999)}]{Patrick1999}%
  \BibitemOpen
  \bibfield  {author} {\bibinfo {author} {\bibfnamefont {J.}~\bibnamefont
  {Brinckmann}}\ and\ \bibinfo {author} {\bibfnamefont {P.~A.}\ \bibnamefont
  {Lee}},\ }\href@noop {} {\bibfield  {journal} {\bibinfo  {journal} {Phys.
  Rev. Lett.}\ }\textbf {\bibinfo {volume} {82}},\ \bibinfo {pages} {2915}
  (\bibinfo {year} {1999})}\BibitemShut {NoStop}%
\bibitem [{\citenamefont {Lee}\ and\ \citenamefont {Shih}(1997)}]{TK1997}%
  \BibitemOpen
  \bibfield  {author} {\bibinfo {author} {\bibfnamefont {T.~K.}\ \bibnamefont
  {Lee}}\ and\ \bibinfo {author} {\bibfnamefont {C.~T.}\ \bibnamefont {Shih}},\
  }\href@noop {} {\bibfield  {journal} {\bibinfo  {journal} {Phys. Rev. B}\
  }\textbf {\bibinfo {volume} {55}},\ \bibinfo {pages} {5983} (\bibinfo {year}
  {1997})}\BibitemShut {NoStop}%
\bibitem [{\citenamefont {Ishii}\ \emph {et~al.}(2014)\citenamefont {Ishii},
  \citenamefont {Fujita}, \citenamefont {Sasaki}, \citenamefont {Minola},
  \citenamefont {Dellea}, \citenamefont {Mazzoli}, \citenamefont {Kummer},
  \citenamefont {Ghiringhelli}, \citenamefont {Braicovich}, \citenamefont
  {Tohyama}, \citenamefont {Tsutsumi}, \citenamefont {Sato}, \citenamefont
  {Kajimoto}, \citenamefont {Ikeuchi}, \citenamefont {Yamada}, \citenamefont
  {Yoshida}, \citenamefont {Kurooka},\ and\ \citenamefont
  {Mizuki}}]{Ishii2014}%
  \BibitemOpen
  \bibfield  {author} {\bibinfo {author} {\bibfnamefont {K.}~\bibnamefont
  {Ishii}}, \bibinfo {author} {\bibfnamefont {M.}~\bibnamefont {Fujita}},
  \bibinfo {author} {\bibfnamefont {T.}~\bibnamefont {Sasaki}}, \bibinfo
  {author} {\bibfnamefont {M.}~\bibnamefont {Minola}}, \bibinfo {author}
  {\bibfnamefont {G.}~\bibnamefont {Dellea}}, \bibinfo {author} {\bibfnamefont
  {C.}~\bibnamefont {Mazzoli}}, \bibinfo {author} {\bibfnamefont
  {K.}~\bibnamefont {Kummer}}, \bibinfo {author} {\bibfnamefont
  {G.}~\bibnamefont {Ghiringhelli}}, \bibinfo {author} {\bibfnamefont
  {L.}~\bibnamefont {Braicovich}}, \bibinfo {author} {\bibfnamefont
  {T.}~\bibnamefont {Tohyama}}, \bibinfo {author} {\bibfnamefont
  {K.}~\bibnamefont {Tsutsumi}}, \bibinfo {author} {\bibfnamefont
  {K.}~\bibnamefont {Sato}}, \bibinfo {author} {\bibfnamefont {R.}~\bibnamefont
  {Kajimoto}}, \bibinfo {author} {\bibfnamefont {K.}~\bibnamefont {Ikeuchi}},
  \bibinfo {author} {\bibfnamefont {K.}~\bibnamefont {Yamada}}, \bibinfo
  {author} {\bibfnamefont {M.}~\bibnamefont {Yoshida}}, \bibinfo {author}
  {\bibfnamefont {M.}~\bibnamefont {Kurooka}}, \ and\ \bibinfo {author}
  {\bibfnamefont {J.}~\bibnamefont {Mizuki}},\ }\href@noop {} {\bibfield
  {journal} {\bibinfo  {journal} {Nat. Commun.}\ }\textbf {\bibinfo {volume}
  {5}},\ \bibinfo {pages} {3714} (\bibinfo {year} {2014})}\BibitemShut
  {NoStop}%
\bibitem [{\citenamefont {Peng}\ \emph {et~al.}(2015)\citenamefont {Peng},
  \citenamefont {Hashimoto}, \citenamefont {Sala}, \citenamefont {Amorese},
  \citenamefont {Brookes}, \citenamefont {Dellea}, \citenamefont {Lee},
  \citenamefont {Minola}, \citenamefont {Schmitt}, \citenamefont {Yoshida},
  \citenamefont {Zhou}, \citenamefont {Eisaki}, \citenamefont {Devereaux},
  \citenamefont {Shen}, \citenamefont {Braicovich},\ and\ \citenamefont
  {Ghiringhelli}}]{Peng2015}%
  \BibitemOpen
  \bibfield  {author} {\bibinfo {author} {\bibfnamefont {Y.~Y.}\ \bibnamefont
  {Peng}}, \bibinfo {author} {\bibfnamefont {M.}~\bibnamefont {Hashimoto}},
  \bibinfo {author} {\bibfnamefont {M.~M.}\ \bibnamefont {Sala}}, \bibinfo
  {author} {\bibfnamefont {A.}~\bibnamefont {Amorese}}, \bibinfo {author}
  {\bibfnamefont {N.~B.}\ \bibnamefont {Brookes}}, \bibinfo {author}
  {\bibfnamefont {G.}~\bibnamefont {Dellea}}, \bibinfo {author} {\bibfnamefont
  {W.-S.}\ \bibnamefont {Lee}}, \bibinfo {author} {\bibfnamefont
  {M.}~\bibnamefont {Minola}}, \bibinfo {author} {\bibfnamefont
  {T.}~\bibnamefont {Schmitt}}, \bibinfo {author} {\bibfnamefont
  {Y.}~\bibnamefont {Yoshida}}, \bibinfo {author} {\bibfnamefont {K.-J.}\
  \bibnamefont {Zhou}}, \bibinfo {author} {\bibfnamefont {H.}~\bibnamefont
  {Eisaki}}, \bibinfo {author} {\bibfnamefont {T.~P.}\ \bibnamefont
  {Devereaux}}, \bibinfo {author} {\bibfnamefont {Z.-X.}\ \bibnamefont {Shen}},
  \bibinfo {author} {\bibfnamefont {L.}~\bibnamefont {Braicovich}}, \ and\
  \bibinfo {author} {\bibfnamefont {G.}~\bibnamefont {Ghiringhelli}},\
  }\href@noop {} {\bibfield  {journal} {\bibinfo  {journal} {Phys. Rev. B}\
  }\textbf {\bibinfo {volume} {92}},\ \bibinfo {pages} {064517} (\bibinfo
  {year} {2015})}\BibitemShut {NoStop}%
\bibitem [{\citenamefont {Varma}\ \emph {et~al.}(1989)\citenamefont {Varma},
  \citenamefont {Littlewood}, \citenamefont {Schmitt-Rink}, \citenamefont
  {Abrahams},\ and\ \citenamefont {Ruckenstein}}]{Varma1989}%
  \BibitemOpen
  \bibfield  {author} {\bibinfo {author} {\bibfnamefont {C.~M.}\ \bibnamefont
  {Varma}}, \bibinfo {author} {\bibfnamefont {P.~B.}\ \bibnamefont
  {Littlewood}}, \bibinfo {author} {\bibfnamefont {S.}~\bibnamefont
  {Schmitt-Rink}}, \bibinfo {author} {\bibfnamefont {E.}~\bibnamefont
  {Abrahams}}, \ and\ \bibinfo {author} {\bibfnamefont {a.~E.}\ \bibnamefont
  {Ruckenstein}},\ }\href@noop {} {\bibfield  {journal} {\bibinfo  {journal}
  {Phys. Rev. Lett.}\ }\textbf {\bibinfo {volume} {63}},\ \bibinfo {pages}
  {1996} (\bibinfo {year} {1989})}\BibitemShut {NoStop}%
\end{thebibliography}%

\end{document}